\documentclass[pdflatex,sn-basic]{sn-jnl}

\usepackage{graphicx}%
\usepackage{multirow}%
\usepackage{amsmath,amssymb,amsfonts}%
\usepackage{amsthm}%
\usepackage{mathrsfs}%
\usepackage[title]{appendix}%
\usepackage{xcolor}%
\usepackage{textcomp}%
\usepackage{manyfoot}%
\usepackage{booktabs}%
\usepackage{listings}%
\usepackage{float}%
\usepackage{tablefootnote}%
\hypersetup{bookmarksdepth=3}

\begin{document}

\title[Methodological Harness in Agentic Software Engineering]{Methodological Harness in Agentic Software Engineering: An Empirical Study on Mining Software Repositories}

\author*[1]{\fnm{Jessica} \sur{D\'iaz}}\email{yesica.diaz@upm.es}
\author[1]{\fnm{Jorge} \sur{P\'erez}}\email{jorgeenrique.perez@upm.es}
\author[1]{\fnm{Sergio} \sur{Gil-Borr\'as}}\email{sergio.gil@upm.es}

\affil*[1]{\orgname{Universidad Polit\'ecnica de Madrid}, \orgaddress{\city{Madrid}, \country{Spain}}}

\abstract{\textbf{Context.} Agentic software engineering requires mechanisms that enable teams to coordinate and govern the work of agents. The framework motivating this study proposes a \textit{methodological harness} composed of eight mechanisms: context engineering, persistent shared knowledge, executable specifications, N-version mindset and parallel agents, normative specifications, structured consultation, evidence-based acceptance and graduated autonomy, which are realized through certain \textit{artifacts} such as rule or context files, specifications, and architectural decisions records. This framework had not yet been empirically validated at repository scale.
\textbf{Objective.} This article analyzes to what extent and in what ways this harness is observable in repositories with agentic activity through two research questions: RQ1 characterizes the adoption of the harness mechanisms through the prevalence, breadth, co-occurrence and temporal evolution of their artifacts; RQ2 examines rule files—one of the main observable artifacts and a persistent source of instructions for the agent—to determine to what extent their content corresponds to the mechanisms that the framework proposes.
\textbf{Method.} Over an existing dataset (AIDev) comprising 116,211 GitHub repositories, we analyzed 5,435 using a design stratified by visibility, i.e., stars indicating popularity and/or reputation. For RQ1, we detected artifacts of the seven observable mechanisms and their introduction dates, and quantified them. For RQ2, we conducted qualitative coding of rule files from 150 repositories.
\textbf{Results.} The population prevalence of at least one mechanism is 21.7\,\%, compared with 65.3\,\% in the most visible repositories. Configurations involving multiple mechanisms are rare. Among repositories that already have a rule file, its content almost always orients the agent and states some norm, whereas persistent shared knowledge, executable specifications, structured consultation and graduated autonomy appear only in a minority.
\textbf{Conclusions.} The harness is empirically observable, but mainly through isolated mechanisms, not as the integrated system that the framework postulates.
}


\maketitle

\section{Introduction}\label{sec:introduction}

Between 2024 and 2026, autonomous coding agents moved from research demonstrations to production use  \citep{he2025llmagents,ahmed2025,amalfitano2026,li2026teammates,liu2026}. The impact on team performance, however, has not been what one might expect given the models’ capabilities. The report by Faros AI, covering more than 10,000 developers across 1,255 teams, links the adoption of generative AI with a 21\,\% increase in tasks completed and a 98\,\% increase in merge requests accepted, but also with a 91\,\% increase in review time and a 9\,\% increase in defects per developer  \citep{farosai2025paradox}. The DORA reports describe a comparable tension and summarize it by stating that AI does not fix a team, but rather amplifies what is already there \citep{dora2024,dora2025}. In the same vein, a randomized controlled trial conducted by METR found that experienced developers took 19\,\% longer to complete tasks in their own repositories while believing they had been about 20\,\% faster  \citep{metr2025}.

The conceptual work preceding this study  \citep{diaz2026sdd} proposes that the limiting factor is not the model capability  but the absorption capacity of the socio-technical model surrounding it, and that what enables this capacity to be absorbed is a \textit{harness}, defined as the set of technical and methodological mechanisms that surround the agent and the team and translate the agent capability into results for the team. The term originates in gray literature as the infrastructure that surrounds a model, such that one could say that \textit{Agent = Model + Harness}  \citep{fowler2026harness,langchain2026harness}, and that work \citep{diaz2026sdd} extends this to the team level and links it to specification-driven development (SDD). The \textit{methodological harness} framework was constructed, deliberately and explicitly, on the basis of gray literature, following the multivocal review protocol of  \citet{GAROUSI:2019}, and is presented as a first step toward academic-industrial consensus rather than as a validated theory. This framework proposes a \textbf{methodological harness} comprising eight mechanisms: (H1) context engineering, (H2) persistent shared knowledge, (H3) executable specifications, (H4)  N-version mindset and parallel agents, (H5) normative specifications, (H6) structured consultation, (H7) evidence-based acceptance and (H8) graduated autonomy. These mechanisms are materialized through certain \textit{artifacts}, such as rule or context files (e.g. AGENTS.md), specifications, and architectural decisions records (ADR), which become persistent sources of instructions for agents. The framework itself leaves open a question that no peer-reviewed study has answered: do these mechanisms exist in practice, and to what extent? 

The natural temptation, given the findings of the industry reports cited above, is to ask whether the harness \textit{works}, that is, whether the repositories that adopt it integrate more frequently, review code more quickly, or reduce the number of code approvals that lack human oversight. This study deliberately does \textbf{not} ask that question, and the reason is not one of scope but of order. This study examines the extent to which the proposed methodological harness for agentic software engineering is observable in real repositories, both in terms of its adoption and the content through which it is materialized. The research questions are:

\begin{itemize}
  \item \textbf{RQ1 --- Adoption.} To what extent are the mechanisms of the methodological harness adopted in repositories involved in agentic development? RQ1 aims to identify the artifacts that materialize the harness, defining the following analysis dimensions: their \textit{prevalence} (how many repositories); \textit{breadth} (how many mechanisms per repository); \textit{co-occurrence} (which ones appear together); \textit{visibility gradient} (how adoption varies with repository visibility); and \textit{temporal evolution} (when they are adopted). These are not additional research questions, but dimensions of the same question.
  
  \item \textbf{RQ2 --- Alignment with the framework.} What do  rule files (the main observable artifacts and persistent source of instructions for  agents) in the repositories contain, and to what extent does that content correspond to the mechanisms postulated by the methodological harness framework?
\end{itemize}

RQ1 uses a deductive operationalization of the framework to identify its mechanisms on a large scale. RQ2 also uses deductive categories (the eight mechanisms of the harness) but allows the data to contradict or go beyond the taxonomy, including the inductive coding of content that does not fit, which this article does not address. 


To answer these questions, we have conducted a descriptive empirical study on an existing dataset called AIDev  \citep{aidev2025dataset} which contains information on 116,211 GitHub repositories with \textit{pull requests} (PRs) created by agents. For \textbf{RQ1} we combined a census of the 2,807 repositories in the most visible stratum (repositories with more than 100 popularity or reputation stars) with a \textit{stratified sample divided into visibility bands} comprising 3,000 repositories from the rest of the population; subsequently, 5,435 repositories out of the 5,807 selected were cloned and analyzed (the difference is due to repositories being renamed, deleted or archived). Stratification by stars is not a sampling technicality but rather the measurement tool for one of the questions. Repository mining studies have long warned that the long tail of personal and single-use projects does not represent a population of team-based software projects  \citep{kalliamvakou2016perils,munaiah2017curating}, and the usual response, namely filtering by popularity, changes the population precisely where one wishes to measure it. The stratified design transforms this bias, which in other studies is a stated threat, into a measurable quantity. Seven of the eight harness mechanisms were operationalized using observable traces across the 5,435 public repositories and their dates of first appearance, but mechanism H4 (N-version mindset and parallel agents) was excluded for leaving no persistent trace in Git. For \textbf{RQ2} we conducted a qualitative analysis of the full content of 285 rules files from 150 repositories, using deductive categories derived from the framework’s mechanisms and independent coding with disagreement adjudication. The study measures presence and materialization; however, it deliberately does not measure effects on development performance, which would be the subject of a subsequent study.




This article is structured as follows. Section~\ref{sec:relacionado} describes agentic software engineering (ASE), the eight harness mechanisms proposed in \citep{diaz2026sdd}, and some brief considerations on software repository mining. Section~\ref{sec:metodo} describes the shared sampling frame and is then organized into two parallel sections corresponding to RQ1 and RQ2: the quantitative study (operationalization and descriptive measures) and the qualitative study (content coding). Section~\ref{sec:resultados} follows the same structure, examining observable adoption (RQ1) and correspondence with the framework (RQ2). Section~\ref{sec:amenazas} describes the limitations and threats of the study. Section~\ref{sec:discusion} interprets the results in light of the framework from which they are derived, including those parts of the framework that these data do not support. Finally, Section~\ref{sec:conclusiones} summarizes the conclusions and outlines future research directions.

\section{Background}\label{sec:relacionado}

\subsection{Agentic Software Engineering} \label{sec:rel-resultado}

Recent literature documents the transition of autonomous coding agents from research demonstrations to production use  \citep{he2025llmagents,ahmed2025,amalfitano2026,li2026teammates,liu2026}, while a series of industry reports assesses their impact on teams that have already adopted them \citep{farosai2025paradox,dora2024,dora2025,metr2025}. These reports suggest that while the adoption of general AI can significantly increase individual productivity, the resulting improvements do not necessarily translate into better outcomes for the team when governance, review, verification and coordination capabilities remain unchanged. Taken together, these findings suggest that the limiting factor is not the model’s capability, but rather the absorptive capacity of the socio-technical model surrounding it. However, the reviewed literature does not establish whether the mechanisms enabling such absorption exist in practice, nor to what extent or what form such adoption actually takes. The study presented here logically precedes all of them: it measures the precondition ---presence and content--- that impact studies take for granted.

\subsection{Harnessing Engineering} \label{sec:rel-arnes}

The notion of a \emph{harness} has emerged in the agentic software engineering literature to denote the infrastructure surrounding the underlying language model, often summarized as \emph{Agent = Model + Harness} \citep{fowler2026harness,langchain2026harness}. In this view, the model alone is only a text-to-text component; the technical harness turns it into an engineering actor by providing an orchestration loop, tools, context management, persistent state, guardrails and permissions, and verification loops. These mechanisms determine what the agent can observe and modify, what information is available in its context, which actions require approval, and how generated changes can be checked and corrected before they are presented to a human.

Building on this technical notion, in a previous work, we  distinguished a second layer, the \emph{methodological harness} \citep{diaz2026sdd}. The distinction is one of level and purpose. Whereas the technical harness primarily governs the execution of one agent in one session, the methodological harness addresses properties that arise at team level, including reproducibility, auditability, transfer of knowledge across sessions and agents, coordination, and the growing review burden created by high-throughput agentic development. Its components are therefore not mainly software around the model, but shared practices and artifacts owned by the team, with specifications acting as the central coordination substrate. The framework further argues that the two layers differ in durability: technical scaffolding may depreciate as model capabilities evolve, whereas methodological artifacts accumulate the team's intent, norms, decisions, and evidence over time.

\begin{figure}[t]
    \centering
    \includegraphics[width=\textwidth]{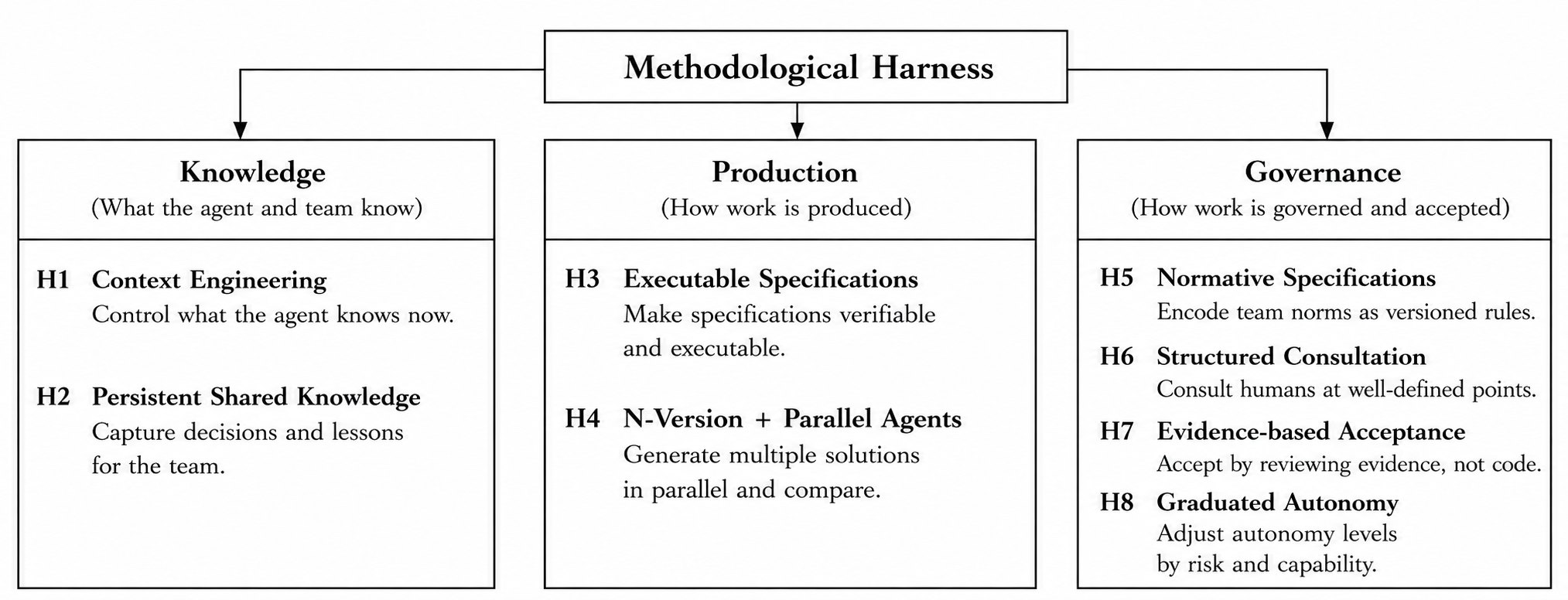}
    \caption{The methodological harness. The eight mechanisms proposed by \citet{diaz2026sdd} are organized into three complementary functions: knowledge management, production support, and governance. Together, they provide team-level infrastructure for scalable human--agent collaboration.}
    \label{fig:mharnes-emse}
\end{figure}

Figure~\ref{fig:mharnes-emse} summarizes the eight mechanisms proposed for the methodological harness and groups them into three complementary functions: knowledge management, production support, and governance.

The eight mechanisms operationalize this team-level infrastructure as follows. \textbf{H1, context engineering}, treats the information environment in which agents operate as a shared, versioned team asset rather than a private per-developer configuration. \textbf{H2, persistent shared knowledge}, preserves decisions, rationale, summaries, and discoveries across sessions, agents, tools, and human participants, compensating for the stateless nature of individual model sessions. \textbf{H3, executable specifications}, requires specifications to be consumable by agents and connected to automated checks through testable acceptance criteria, enforceable constraints, and explicit decisions. \textbf{H4, N-version mindset and parallel agents}, treats inexpensive parallel generation of candidate solutions as a normal mode of work and shifts human effort toward comparison, evaluation, and selection; the framework associates this mode with isolated working trees so that parallel agents do not interfere at file-system level.

The remaining mechanisms concern durable rules and human control. \textbf{H5, normative specifications}, makes standing conventions, patterns, and prohibitions explicit, versioned, and reviewable parts of the system specification rather than tacit knowledge held by senior developers. \textbf{H6, structured consultation}, replaces continuous supervision with scoped intervention: when an agent encounters a decision it cannot or should not resolve, it suspends execution and raises a consultation containing the question, relevant options, specification context, and consequences, with the resolution recorded for future work. \textbf{H7, evidence-based acceptance}, shifts the unit of completion from a merged pull request to a structured evidence bundle demonstrating conformance with the originating specification, including functional evidence, verification results, maintainability checks, rationale, and audit links. Finally, \textbf{H8, graduated autonomy}, treats autonomy as an explicit and revisable property of classes of work rather than as a global on/off setting, allowing teams to calibrate when agents may act autonomously and when human planning, approval, consultation, or review is required.

These mechanisms are intended to operate as a coherent system rather than as independent practices. Executable specifications make evidence-based acceptance possible; parallel generation increases the need for explicit evaluation and autonomy boundaries; consultation and review produce decisions that can become normative guidance; and persistent shared knowledge carries those decisions across subsequent sessions and agents. In this sense, the methodological harness proposes a cumulative organizational layer around agentic development, with specifications linking context, production, verification, and governance.

Importantly, \citet{diaz2026sdd} present this taxonomy as a conceptual synthesis grounded partly in practitioner and gray literature, developed following a multivocal-review perspective \citep{GAROUSI:2019}, rather than as an empirically validated theory of team performance. The present study therefore does not assume that the eight mechanisms are effective or commonly adopted. Instead, it uses the taxonomy as the construct to be operationalized against repository evidence, measuring which mechanisms leave observable traces in real projects, how frequently they occur, and where the empirical data do not support the framework's expectations.

To the best of our knowledge, this study is the first empirical test of this taxonomy against a corpus of repositories: it does not simply accept it, but tests it against the data, and explicitly reports where the data do not support it  (Section~\ref{sec:discusion}).

\subsection{Repository mining: population, bias and protocol} \label{sec:rel-mineria}

Repository mining has been highlighting for a decade that  \texttt{GitHub} is not a homogeneous population of team projects: it contains a long tail of personal repositories, templates and one-off experiments that a popularity filter excludes, at the cost of skewing the resulting population  \citep{kalliamvakou2016perils,munaiah2017curating}. This study addresses that concern through design rather than filtering (that is, by stratifying by visibility and reweighting to the full population, Section~\ref{sec:metodo-diseno}) so that popularity bias becomes a measured quantity rather than a stated but unquantified threat.

In terms of protocol, this study deliberately departs from the norm of formal pre-registration that has gained ground in the empirical sciences following the replication crisis  \citep{nosek2018preregistration}: as it contains no outcome variables nor tests any association hypotheses  (Section~\ref{sec:metodo-diseno}), there is no quantitative prediction to protect from post-hoc adjustment of the data. Instead, equivalent safeguards are applied where there is a risk of adjustment: the codebook is frozen before coding begins, the reliability threshold  $\kappa \geq 0{.}60$ is set in advance, and the code and every coding label are published in full (Section~\ref{sec:metodo-cualitativo}, Section~\ref{sec:metodo-repro}). Formal pre-registration will be mandatory as soon as the study incorporates outcome variables, and that is precisely the dividing line between this work and that which is stated as pending (see Section~\ref{sec:discusion}).

\section{Research Methodology}\label{sec:metodo}

This section describes an empirical study of software repository mining on AIDev \citep{aidev2025dataset}. This study is  \emph{descriptive}. Its aim is to establish to what extent repositories with agentic activity  (repositories with PR ---\emph{pull request}--- executed by agents) adopt the mechanisms of the harness and how they implement them, before asking whether this adoption produces any effect. The distinction is not merely rhetorical: it determines what is measured and what is not. Two commitments accompany this approach:

\begin{itemize}
  \item \emph{No null-hypothesis testing.} Given the descriptive nature of the study, we formulate research questions rather than directional hypothesis, and report descriptive estimates with their uncertainty instead of null hypothesis significance testing (NHST). 
  \item \emph{Every published figure is traceable } to a named script and a fixed version of the dataset.
\end{itemize}

\subsection{Study design}\label{sec:metodo-diseno}

We first conducted a pilot study on 10 repositories (677 PRs, 2,955 comments) to validate the instrument before scaling up. We then conducted the main study. This study combines, on the one hand, quantitative data mining of repositories at scale, and, on the other, qualitative content analysis:

\begin{itemize}
  \item \textbf{Quantitative component.} This focuses on the observable adoption of the mechanisms of the harness (RQ1). Seven of the eight mechanisms are operationalized through observable traces in 5,435 public repositories, which were cloned, together with the dates of first appearance of these traces; H4 (N-version mindset and parallel agents) was excluded as it does not leave a persistent trace in Git. To this end, we wrote the code of automatic artifact detectors (data mining scripts written in Python that detect the existence of certain file paths or the presence of certain lexical patterns), and calculated the prevalence, breadth, co-occurrence and temporal evolution of the mechanisms.
  
  \item \textbf{Qualitative component.} This examines whether the content of rule files actually materializes the constructs postulated by the framework (RQ2). A qualitative analysis is conducted on the full content of 285 rule files from 150 repositories, using deductive categories derived from the framework’s mechanisms and independent coding with disagreement adjudication. Additionally, a qualitative analysis is conducted on the comments present in PRs from repositories in the most visible stratum (P).
\end{itemize}

\begin{figure}[h]
\centering
\includegraphics[width=\textwidth]{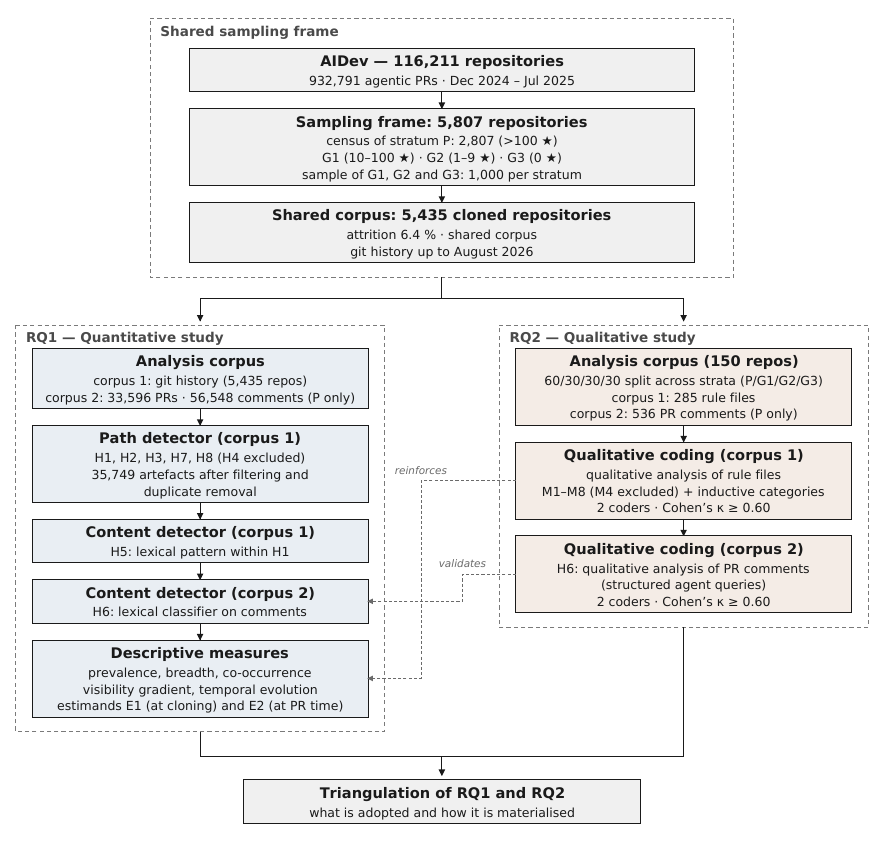}
\caption{Study design. Both components share the same sampling frame and unit of analysis (the repository); RQ1 measures presence using automatic detectors and RQ2 examines the content of the artifacts through human coding.}\label{fig:diseno}
\end{figure}

Figure~\ref{fig:diseno} summarizes this design: both components share the sample frame and the unit of analysis (the repository); RQ1 measures presence of the artifacts using automatic detectors, and RQ2 examines the content of the artifacts through human coding.

\subsubsection{Data source}\label{sec:metodo-datos}

The sample frame is drawn from AIDev  \citep{aidev2025dataset}, a dataset of PRs created by autonomous coding agents on GitHub. Data collection began in December 2024 and PR coverage ended on July 30, 2025, so AIDev’s observation window is
approximately 7 months. The last declared update to the dataset hosted on Hugging Face was to the metadata (not new data) on May 10, 2026. For this study, revision  \texttt{68ed5f4b80d27a9e057fc57567f38bd322ac73ec} is used, and all counts in this article are recalculated based on it (script \texttt{src/ingest/cache\_tables.py}; manifest in \texttt{data/step\_zero\_report.json}).

The end date (end of AIDev PRs coverage) requires clarification. This cut-off date (July 30, 2025) determines which repositories are included in our sample, not how long their artifacts are observed. The indicators in this study are measured based on the \textit{Git history} of each cloned repository, which therefore extends to the cloning date (August 2026), not over AIDev’s PRs. Consequently, what is derived from the cut-off is a selection: the repositories observed, and which we have cloned, are those that already had agent activity prior to July 30, 2025, that is, early adopters of agents. It is plausible that those who adopt agents early will also adopt the harness early, so the prevalence estimates given here probably \textit{overestimate} those of the general population of repositories with agentic activity. This is declared as a transferability limit and is not corrected, because correcting it would require assuming the magnitude of the bias is known.


It is also worth noting that AIDev is organized into two tiers. The \textit{full} tier covers 116,211 repositories and 932,791 agentic PRs, containing only repository and PR metadata; the \textit{popular} tier contains 2,807 repositories with more than 100 stars and includes PR comments, reviews and timelines (a full breakdown by tier is provided in Appendix~\ref{sec:apx-aidev}, Table~\ref{tab:apx-aidev} and Table~\ref{tab:apx-comentarios}).

\subsubsection{Population and sample design}\label{sec:metodo-muestreo}

The population consists of 116,211 repositories. The design consists of a census of the \textit{popular} stratum (2,807 repositories with  $>$100 stars) plus a stratified sample by visibility bands comprising 3,000 repositories from the remainder of the population  (see Table~\ref{tab:estratos}): 1,000 repositories from stratum G1 (repositories with  $\geq$10 stars); 1,000 from stratum G2 (repositories with 1--9 stars); and 1,000 from stratum G3 (repositories with 0 stars). Note that each of the samples in P, G1, G2 and G3 is statistically significant in representing their respective strata and the overall reweighted population, assuming the standard 95\,\% confidence level ($Z$ = 1.96) and maximum variance ($p$ = 0.5) used in research. Both the quantitative study (RQ1, see Section~\ref{sec:metodo-rq1}) and the qualitative study  (RQ2, see Section~\ref{sec:metodo-cualitativo}) are based on this same sampling framework.

Stratification by star rating is not a sampling technicality: it is the instrument used to measure one of the questions. A simple random sample of 3,000 repositories from the entire population would have assigned around 2,421 cases to the 0-star band and barely 95 to the $\geq$10-star band, leaving the comparison between bands (that is, the visibility gradient) unresolvable precisely where it was intended to be measured. The design transforms popularity bias, which in other data mining studies is a threat that is acknowledged, into a quantity that is measured.

\begin{table}[h]
\caption{: Sampling design. Stratum P is defined as belonging to the repository table of the \textit{popular} tier, without recalculating stars, so that the definition is reproducible for the specified revision. ``1 PR'' is the proportion of the stratum with exactly one PR agentic. ``\%'' is the weight of each stratum relative to the total population (116,211).}\label{tab:estratos}
\footnotesize
\begin{tabular*}{\textwidth}{@{\extracolsep\fill}llrrrrrr@{\extracolsep\fill}}
\toprule
Stratum & Definition & Population & Sample & \% & Weight & No license & 1 PR  \\
\midrule
P  & \emph{Popular} tier ($>$100 stars) & 2,807   & 2,807 & 2.4\,\%  & 1.00        & 6.8\,\% & 43.9\,\%  \\
G1 & Complement, $\geq$10 stars     & 3,665   & 1,000 & 3.2\,\%  & 3.67        & 19.0\,\% & 45.6\,\%  \\
G2 & Complement, 1--9 stars         & 15,941  & 1,000 & 13.7\,\% & 15.94       & 49.7\,\% & 42.4\,\% \\
G3 & Complement, 0 stars            & 93,798  & 1,000 & 80.7\,\% & 93.80    & 71.7\,\%    & 48.4\,\%  \\
\midrule
   & Total (reweighted)                 & 116,211 & 5,807 & 100.0\,\% &          &             & 47.4\,\%      \\
\botrule
\end{tabular*}
\footnotesize\emph{Source: src/mine/select\_sample.py \texttt{->} data/sample\_frame.csv, data/sample\_frame\_report.json.}
\end{table}


Not all items in the population are team software projects. 47.4\,\% of repositories have exactly one agentic PR, 80.7\,\% have 0 stars and 71.7\,\% of the G3 stratum do not declare a license  (see Table~\ref{tab:estratos}): this is the long tail of personal experiments, templates and single-use repositories documented by  \citet{kalliamvakou2016perils} and \citet{munaiah2017curating}. The population is not filtered. Instead, every main figure is reported twice: for 100\,\% of the population (including long-tail noise) and for a restricted version (only repositories with \textit{$\ge 2$ agentic PRs}, filtering out one-off contributions). If the total and restricted figures match, this means that the phenomenon under study is stable and the ‘long tail’ of disposable projects did not affect the result. If the figures diverge, the difference is not an error, but an empirical finding. It reveals exactly where the practice is found; that is, whether agentic behavior is a widespread standard or whether it is a practice exclusive to structured/popular projects.

The sampling is \textit{deterministic} and reproducible: within each band, the repositories are sorted by  \texttt{md5(id~$\|$~semilla)} and the first 1,000 are selected, using the seed \texttt{001v2-20260817} recorded in the code. Anyone who runs  \texttt{src/mine/select\_sample.py} on the specified commit will obtain exactly the same sample. The population counts in Table~\ref{tab:estratos} are produced by that same script and match those stated in the study specification.

\subsubsection{Attrition}\label{sec:metodo-desercion}

Every repository in the study ends up in exactly one of these three states ---\emph{cloned}, \emph{API backup } o \emph{unavailable}--- and the complete record is published. When cloning fails, the GitHub API is queried to distinguish between three situations that a single counter would conflate: (i) a renaming (the repository remains within the scope under a different name; the cloning is retried with the resolved name, and if the retry is successful, the repository is counted as \textit{cloned}, not as an API backup), (ii) an archived repository for which cloning is not retried, and (iii) a deletion or change to private status. The \textit{API backup} category therefore groups together cases where the API confirmed the repository’s status but a usable clone was not obtained.

Attrition is reported \textit{by stratum rather than in aggregate}, because it increases as one moves down through the visibility bands, and this variation is informative in itself: 2.7\,\% in P, 3.5\,\% in G1, 11.8\,\% in G2 and 14.4\,\% in G3 (see Table~\ref{tab:disposicion} en la Section~\ref{sec:res-muestra}). This progression necessitates a \textit{warning regarding differential attrition}. In G3, one in seven repositories in the framework is no longer accessible — either deleted or made private — and the prevalence figures for that stratum are estimated, by construction, on the basis of the remaining repositories (survivors). If the repositories that disappear are disproportionately short-lived experiments, and therefore less well-harnessed, the G3 estimate is biased upward, and the actual gradient would be even more pronounced than that observed. The bias would therefore work against the finding, not in its favor, which makes for a conservative interpretation; however, the direction cannot be verified using data from repositories that no longer exist, and this is stated as such.

This sample framework (5,435 cloned repositories out of 5,807 selected) forms the corpus on which both the quantitative study for RQ1  (see Section~\ref{sec:metodo-rq1}) and the qualitative sample for RQ2 (see Section~\ref{sec:metodo-cualitativo}) are based.

\subsection{Quantitative study: RQ1 — observable adoption}\label{sec:metodo-rq1}

This section describes the instrument used to answer RQ1: artifact extraction, its cleaning and deduplication, operationalization of the harness mechanisms, definition of indicators and two temporal estimands, their aggregation into descriptive measures, instrument validation and the special case of harness H5 (normative specifications), and the instrument’s language coverage.

\subsubsection{Artifact extraction}\label{sec:metodo-mineria}

The extraction is \emph{clone-first}. A bare clone is obtained from each repository (the working tree is completely discarded) with a filter for large objects,

\begin{center}
\texttt{git clone -{}-bare -{}-filter=blob:limit=100k}
\end{center}

\noindent and, on this clone, the following are determined in a single pass and without using up the REST API quota: both the presence and the commit dates of the artifacts that implement the  harness’s mechanisms. Each indicator is obtained via a detector: the procedure, implemented in the mining script, which determines, based on the contents of the repository, whether a mechanism is present. Most detectors involve a deterministic file-path match  (Section~\ref{sec:metodo-mecanismos} lists the detector for each mechanism, such as rule files, specifications and ADRs, or more specifically  \texttt{CLAUDE.md}, \texttt{AGENTS.md} o \texttt{CODEOWNERS}). In addition, we use two more sophisticated detectors that operate on textual content rather than on the path: the H5 detector, which counts modal clauses, and the language detector, which classifies the dominant language of each file  (both detailed in  Section~\ref{sec:metodo-idioma}). Here, the GitHub API is used only as a fallback when cloning fails (see Section~\ref{sec:metodo-desercion}). The date of first appearance for each artifact is obtained from \texttt{git log -{}-diff-filter=A -{}-format=\%aI -{}- <path>}, taking the oldest entry. Presence alone is never sufficient: the two estimands in  Section~\ref{sec:metodo-estimandos} and the adoption curve rely on these dates. The script \texttt{src/mine/mine\_harness.py} performs this extraction; its output by stratum is stored in \texttt{data/harness\_\{P,G1,G2,G3\}\_raw.json}.

The clones are discarded after extraction. Retaining all 5,807 would have required around  100--200\,GB; discarding them reduces the disk space required due to the concurrent processing of the repositories, rather than the sample size. The text of the \textit{rule files} is saved before the clone is discarded, so that the qualitative phase does not have to clone them again. Progress is recorded per repository, making the execution resumable and idempotent.

\subsubsection{Deduplication (data cleansing)}\label{sec:metodo-dedup}

The corpus on which the detectors operate is the file tree and the \texttt{git} history of each cloned repository, not the artifacts extracted from it: the figures in this section are the output of the data mining carried out on that corpus, not a measure of its size. Once a repository has been cloned, the artifacts that operationalize the mechanisms of the harness are extracted. A single file with multiple names counts as one artifact, not several. Before counting, the following are excluded, in this order: (i) symbolic links —\texttt{git mode 120000}—; (ii) artifacts without a resolvable first appearance date in the \texttt{git} history, which is required for the temporal estimands in Section~\ref{sec:metodo-estimandos}; and (iii)~byte-for-byte duplicates  ---same \textit{blob} SHA within the repository  ---, retaining the path entered previously as the canonical one. All three cases are common; consequently, out of the 5,435 cloned repositories, 310 symbolic links, 20 artifacts without a resolvable date and 737 exact duplicates were discarded (1,067 in total). Apart from this exclusion, the stub-files \footnote{A stub file is a rules file  (H1 ---\texttt{AGENTS.md}, \texttt{CLAUDE.md}, \texttt{.cursorrules}, etc.--- which exists in the repository but has virtually no content: fewer than 200 bytes, and no modal clauses (\texttt{MUST}, \texttt{NEVER}, \texttt{do not}\ldots). In practice, it is a file that merely ``points'' to something else  (for example, a line such as ``see \texttt{docs/agents.md} for the full rules,'' or an almost empty file created by a template) rather than containing the actual rules. It is excluded from the count because, otherwise, a repository with an empty  \texttt{CLAUDE.md} file would count as an ``H1 adopter'' in the same way as one with a substantial rules file: this artificially inflates the prevalence without there being any real implementation behind it.}
 are not removed from the count: they are flagged (240 of them), and this flag is used solely for the prevalence of H1, not for the raw count of artifacts, because treating them as absent or present is a debatable judgment, the sensitivity of which is measured by separately (Section~\ref{sec:res-sensibilidad}, Table~\ref{tab:sens-stubs}).

\emph{The count is per mechanism, not per path.} A repository containing sixteen files  \texttt{.cursor/rules/*.mdc} reflects a tool convention, not sixteen times more harness; counting by path would make it the most heavily harnessed repository in the sample. Every key figure is also reported both with and without the stub-files filter, the only one of the four data-cleaning criteria that involves a debatable judgment.

\subsubsection{Operationalization of harness mechanisms}\label{sec:metodo-mecanismos}

The conceptual framework postulates eight mechanisms  (see H1--H8 in Section~\ref{sec:rel-arnes}). Table~\ref{tab:mecanismos} lists those that leave an observable trace in a repository and how they are detected by means of a set of patterns. Two asymmetries are described below which should be noted:

\paragraph{Asymmetry 1: H4 (N-version mindset and parallel agents) leaves no trace on the server}
H4 leaves no server-side trace, and not through an oversight of the instrument: the harnesses of the agentic tools themselves (Claude Code, Codex, OpenCode, among others) implement it internally, as session or worktree isolation, rather than as an artifact that the clone can retain  (e.g. native \emph{git worktrees} in Claude Code \citep{claudecode2026worktrees}, cloud environments or \textit{worktrees} in Codex  \citep{openai2026codexcloud} and, in a less integrated manner, independent OpenCode processes on the same repository  \citep{opencode2026agents}). It is therefore excluded: it is not modeled, approximated or reported. Consequently, seven of the eight harness mechanisms are studied.

\paragraph{Asymmetry 2: H6 (structured consultation) is only observable in stratum P (popular tier)}
\emph{H6} leaves no trace in the file tree either, but for a different reason: the only observable record that an agent suspended its execution to ask a bounded question is the text of PR and issue comments. It is therefore the one mechanism detected not by path but by a classifier over lexical patterns (\texttt{src/classify/consultation\_full.py}), and the only one observable in stratum P alone, because AIDev distributes PR comments for that tier only. Consequently, the \textit{breadth index} ranges $0-6$ in P and $0-5$ in the G strata, so any comparison between strata is based on the \textit{five common mechanisms} (H1, H2, H3, H7, H8). 

For the 2,807 repositories in stratum P and their 33,596 agentic PRs, the detector examines 99,225 text units (PR bodies, comments, and review comments; the breakdown by surface is in Appendix~\ref{sec:apx-aidev}, Table~\ref{tab:apx-comentarios}), of which 56,559 were authored by an agent. Because a lexical classifier can miss what it does not name, its output is validated against human coding, and  whether H6 enters the quantitative analysis depends on that validation described in Section~\ref{sec:metodo-medida-pi2}. 

The classifier reads two lexical signals in each agent-authored text and applies three exclusion rules. The \emph{question} signal is an explicit request for a decision (\texttt{which option}, \texttt{should I}, \texttt{do you prefer}, \texttt{please confirm}\ldots), or a question mark in a text that also declares itself blocked. The \emph{options} signal is two or more distinct enumerated alternatives (\texttt{option 1}, \texttt{option A}, a numbered or lettered list, \texttt{either\ldots or}), counted after stripping checklist lines so that task checkboxes are not read as choices. The exclusions discard text that answers a human (\texttt{great question}, \texttt{you are right}), text that reports a finished task or a review, and pull-request template boilerplate. These two signals define the five mutually exclusive blocks of Table~\ref{tab:h6-cajones}: both signals together with at least one corroborating cue---the text cites a specification, declares itself blocked, or states a trade-off---is \emph{strict detection}; both signals without any such cue is \emph{lax detection}, and these two blocks are the detector's positives. \emph{Question, without options} and \emph{Options, without question} carry one signal each, and \emph{No signal} carries neither. Texts carrying both signals but falling under an exclusion rule---11 of them---belong to no block, which is why the blocks add up to 56,548 rather than to the 56,559 agent-authored texts. The sampling design is fixed before coding: blocks whose population is manageable are read in their entirety (census), and the two that are not are capped at 150---a size chosen so that the upper bound at the 95\,\% confidence level in the worst case (zero confirmed positives) does not exceed approximately 2\,\%. The population of each block and the outcome of the human validation are reported in Table~\ref{tab:h6-cajones} (Section~\ref{sec:res-h6-validacion}). 

\begin{table}[h]
\caption{Operationalization. The list of patterns forms part of the instrument.}\label{tab:mecanismos}
\footnotesize
\begin{tabular*}{\textwidth}{@{\extracolsep\fill}llp{0.52\textwidth}@{\extracolsep\fill}}
\toprule
& Mechanism & Detector \\
\midrule
H1 & Context engineering & \texttt{AGENTS.md}, \texttt{CLAUDE.md},
     \texttt{agent.md}, \texttt{GEMINI.md}, \texttt{.cursorrules},
     \texttt{.cursor/rules/**}, \texttt{.windsurfrules}, \texttt{.clinerules},
     \texttt{.github/copilot-instructions.md} \\
H2 & Persistent shared knowledge & Architecture Decision Record (ADR) in \texttt{docs/adr/**},
     \texttt{docs/decisions/**}, \texttt{adr/**}; \texttt{.specify/memory/**} \\
H3 & Executable specifications & \texttt{specs/**},
     \texttt{.specify/specs/**}, Gherkin \texttt{**/*.feature},
     \texttt{openapi.\{yaml,json\}} \\
H4 &  N-version mindset and parallel agents & \emph{No server-side trace: implemented within the harness, not as a repository artifact. Excluded from the study} \\
H5 & Normative specifications & Lexical pattern over the text of the H1 rule files
     (\texttt{src/mine/mine\_harness.py}): modal clauses marking a standing norm
     (\texttt{MUST}, \texttt{NEVER}, \texttt{do not}\ldots).
     Indicator conditioned on H1 (Section~\ref{sec:metodo-h5}) \\
H6 & Structured consultation & Lexical classifier over PR and issue comments (\texttt{src/classify/consultation\_full.py}): agent-authored text posing a bounded question with enumerated options. Only observable in stratum P. 
      \\
H7 & Evidence-based acceptance & \texttt{.github/pull\_request\_template.md} and
     variants \\
H8 & Graduated autonomy & \texttt{CODEOWNERS} \\
\botrule
\end{tabular*}
\footnotesize\emph{Source: src/mine/mine\_harness.py (path patterns for H1--H3, H7, H8; lexical pattern for H5) and src/classify/consultation\_full.py (H6); the patterns are published in full within the scripts themselves.}
\end{table}


\paragraph{Measurement bias} The instrument is designed to be conservative: it is more likely to underestimate the prevalence of the harness than to overestimate it. The detectors are based on canonical paths and lexical patterns, and they do not attempt to detect variants or synonyms. For example, a repository that implements H1 in a file named \texttt{AGENT\_RULES.md} will not be detected, because the detector only looks for the canonical names listed in Table~\ref{tab:mecanismos}. This is a deliberate design choice: the goal is to measure adoption of the harness as defined by \citet{diaz2026sdd}, not to measure adoption of similar mechanisms under different names. The result is that the prevalence figures reported here are lower bounds on the true prevalence of harness-like mechanisms in the population (a threat discussed in Section~\ref{sec:amenazas}). The direction of this \emph{measurement bias} matters: it runs against the authors' own position on the usefulness of the harness, rather than in its favor. This is the preferable direction for an instrument with which its own proponents put their framework to the test, and it is the reason for publishing the complete list of patterns in Table~\ref{tab:mecanismos}: without it, the figure is not interpretable.

\subsubsection{Indicators}\label{sec:metodo-indicadores}

The indicators in this study are adoption indicators; there are no outcome indicators. For each repository and mechanism, the binary presence at the time of cloning  (that is, under the \emph{temporal estimand E1} defined below, in Section~\ref{sec:metodo-estimandos}) and the date of first appearance of the canonical artifact are recorded. From these, the \textit{breadth} is derived as the count of mechanisms present. Added to this is the density of normative clauses per KB in the rule files, and descriptors used solely for disaggregation (stars, language, license, age, volume of agentic PRs, mix of agents). 

\subsubsection{Temporal estimands}\label{sec:metodo-estimandos}

The statement  \emph{``the repository has the mechanism''} can be interpreted in two different ways: evaluated at the time of cloning (E1) or evaluated at the moment each PR was created (E2). The difference between the two is not a minor one: for several mechanisms, the difference between the value of E1 and the value of E2 is more than double  (Section~\ref{sec:res-e2} of the results details the ratios by mechanism in Table~\ref{tab:e2}). For this reason, both interpretations are always reported, never just one:

\begin{description}
  \item[E1 --- prevalence of the practice.] The artifact exists in the repository’s current \texttt {HEAD}. It answers: \emph{how many teams had adopted the mechanism at the time of cloning?} This is the key figure in the prevalence analysis.
  \item[E2 --- coverage of the observed work.] For each PR, the mechanism counts as present if its first appearance date is earlier than or equal to the creation date of that PR. It answers: \emph{what proportion of the recorded agentic work took place under a harness?}
\end{description}

The difference between the two \textit{is} a result: it measures how much adoption has occurred after the observed work (the PR) was carried out. For example, if a repository uploads its \texttt{CLAUDE.md} file in March 2025, a PR opened in January 2025 in that same repository counts as ‘\textit{without} the mechanism’ under E2 because the file did not yet exist, even though at the time of cloning (E1) the repository does have it. An automatic check ensures that this rule is always followed: No E2 indicator can be derived from an artifact that first appeared after the creation of the PR.

\subsubsection{Descriptive measures}\label{sec:metodo-agregacion}

By aggregating and re-weighting the indicators, we obtain the five descriptive measures that address RQ1: \emph{prevalence} (Section~\ref{sec:res-prevalencia}), the proportion of repositories per stratum with the presence indicator active for a mechanism; the \emph{visibility gradient } (Section~\ref{sec:res-forma}), showing how this prevalence varies across the strata P, G1, G2 and G3; the \emph{breadth} (Section~\ref{sec:res-amplitud}), the distribution of the count of mechanisms present per repository; the \emph{co-occurrence} (Section~\ref{sec:res-coocurrencia}), how many repositories simultaneously host each pair of mechanisms; and the \emph{temporal evolution} (Section~\ref{sec:res-adopcion}), the monthly distribution of the dates of first appearance of each mechanism. 

\paragraph{Re-weighting} Any figure that spans more than one stratum is reweighted according to population size, \[ \hat{p} \;=\; \frac{\sum_h N_h\,\hat{p}_h}{\sum_h N_h}, \] where $N_h$ is the population of stratum h. Without it, the 1,000 zero-star repositories in the sample would be weighted as 1,000 rather than the 93,798 they actually represent. The key figure is, nevertheless, the table by stratum, not the aggregate: the comparison between strata is the result, while the aggregate  ---80.7\,\% of which consists of repositories with no stars --- answers a different and less informative question. Both are published, and it is made clear which is which.

\paragraph{Uncertainty} The intervals are $95\,\%$ Wilson score intervals. Several proportions of interest live near zero, where the normal approximation produces negative lower bounds and misleadingly narrow widths; the Wilson interval behaves correctly in that regime.

\subsubsection{Validation of the instrument}\label{sec:metodo-h5}

This section documents three checks applied to the measuring instrument, each motivated by a specific problem detected during the analysis: a hidden dependency between two mechanisms, which necessitated treating one of them (H5) as a property of the other (H1) rather than as an independent mechanism; a saturated pattern that behaved like a constant and was removed from the detector (H8); and the general requirement that every detector must demonstrate that it triggers before one of its zeros is reported as such. These three findings determine how the prevalence and breadth figures reported later must be read (Table~\ref{tab:prevalencia}, Table~\ref{tab:amplitud}).

\paragraph{H5 (Normative specifications)} H5 is operationalized by counting modal clauses \textit{within the rule files}, that is to say, within the H1 artifacts. Upon examining the co-occurrence matrix for the census of stratum P, it was observed that the co-occurrence of H1 and H5 equaled \textit{exactly} the count of H5: there is no record containing H5 but not H1, and by construction there cannot be. H5 is therefore not an independent mechanism but a property of the content of H1. Retaining it as a separate row in the prevalence table would imply an independence that the operationalization excludes, and would inflate the breadth index by counting the same artifact twice. The decision we took was: \textit{H5 is removed from the main table and from the breadth index, and is reported as an indicator conditional on H1}. H5 conditional on H1 also tells us something stronger and more interesting than a marginal prevalence: \textit{of the repositories that have a rules file, what proportion writes norms into it}.

The case of H5 instances a rule that applies to every mechanism in the study: before treating it as an independent mechanism, a check is carried out to ensure that its co-occurrence with any other mechanism does not equal its own count. When it does equal it, there are not two mechanisms, but rather one mechanism counted twice, and this is detected in the co-occurrence matrix.

\paragraph{H8 (Graduated autonomy)} The \texttt{.github/workflows/**} pattern was removed from the H8 detector after it was found in the pilot study that it flagged 88\,\% of PRs, in other words, virtually every active repository has continuous integration, causing the metric to become so saturated that it effectively became a constant. It was agreed as a rule that \textit{any metric with a prevalence of over 85\,\% should be treated as a hidden constant} and either justified or removed. 

\paragraph{Every detector requires a trigger test} A zero is only reported as a measured zero if, in addition, there is a test on a synthetic repository confirming that the detector does indeed trigger when the corresponding artifact is present. Without that prior test, a genuine zero and a detector that simply does not work are indistinguishable.

\subsubsection{Corpus language, and why it is measured}\label{sec:metodo-idioma}

This section analyzes the language in which the rule files (H1) are written. This analysis extends beyond H1 because the H5 detector counts modal clauses within the content of those same files (Section~\ref{sec:metodo-h5}) rather than examining an artifact of its own: a language detection error in the H1 file therefore propagates directly to the H5 count. The H5 detector counts modal clauses with a pattern found exclusively in English. This choice goes unnoticed while the corpus remains English-only, and ceases to be so as soon as it is not: a rule file written in another language does not match anything and is recorded as lacking normative content, indistinguishable from one that genuinely does not have any. That is why it is important to measure what percentage of the corpus is not in English, as this sets an upper limit on the detection error affecting the H5 figures published in the study, and turns a threat that until now has only been theoretical into a concrete figure.

Detection is carried out using \texttt{langdetect} on the text in the rule files, with two precautions which, in this context, are not merely theoretical:

\begin{itemize}
  \item Code is stripped prior to detection. Fenced blocks, paths, URLs, command-line flags, and indented snippets are removed from the text. A Spanish-language \texttt{CLAUDE.md} filled with \texttt{npm run build} and filenames gets classified as English by pure volume of technical ASCII tokens---and the error runs in the direction that makes the problem \textit{smaller than it actually is}---.
  \item Detection is performed at the paragraph level, and \textit{mixed} is a dedicated category. A corpus may consist of English prose with a section in another language, and a file-level label would erase that distinction. An item is marked as mixed when at least 10\,\% of its non-empty paragraphs are detected in a language other than the dominant one. This category is not cosmetic: in the sample, there are 23 mixed items compared to only 9 whose dominant language is not English.
\end{itemize}

The detector initializes with a fixed seed. Without it, it is non-deterministic, and the same file could be classified in two different ways in two separate runs, which would make the figures irreproducible.

Re-detecting the language of the 4,146 raw-mined rule files (H1) (Table~\ref{tab:artefactos-mecanismo}) would be unnecessary work, because a detection error can only occur where the H5 detector counted zero modal clauses. Sensitivity is therefore calculated solely on that subset: the repositories whose H1 artifacts --- excluding repositories whose only H1 is a stub-file, marked but not counted as H1 (Section~\ref{sec:metodo-dedup}) ---  score \textbf{all} zero clauses (248 repositories; the breakdown by stratum is in  Section~\ref{sec:res-prevalencia}). A file written in English with zero modals is a genuine zero; a non-English file is a measurement error, and the published H5 figures do not distinguish between them. Excluding from the denominator those that the tool cannot read is a conservative correction: it does not assert that those repositories do contain standards, only that we cannot assert that they do not.

\subsection{Qualitative study: RQ2 — alignment with the framework}\label{sec:metodo-cualitativo}

This section describes the methodology used to address RQ2 through a qualitative analysis of the full content of the rule files from 150 repositories, using \textit{deductive categories} derived from the framework’s mechanisms and independent coding with disagreement adjudication.

\subsubsection{Protocol} The instruments are: a deductive codebook frozen prior to coding, open-ended inductive coding for content that does not fit, two independent coders with at least a 20\,\% overlap, Cohen's$\kappa$ \citep{cohen1960kappa} calculated on that overlap with a threshold of $\kappa \geq 0{.}60$ \citep{landis1977agreement}, and disagreements adjudicated by a third author, whose adjudicated labels—--and not those of either of the two coders—--constitute the reference pattern.

\subsubsection{Iterative procedure} Coding proceeds in numbered iterations: a block is encoded, and $\kappa$ is calculated based on its overlap. If the threshold is exceeded, any remaining disagreements are adjudicated, and from that point  on, the coders can operate independently. If the threshold is not exceeded, the code book is corrected (the correction is drafted, its version number is incremented, and it is frozen before being applied to new data), and a different block is re-encoded. A category that still fails to reach the threshold after a correction is removed from the study, with one exception: if adjudication attributes the disagreement to a one-time error by the coder (for example, a transcription error) rather than to a genuine difference in interpretation, correcting that error and recalculating $\kappa$, is legitimate, and the category is retained if the corrected $\kappa$ exceeds the threshold. The coding instrument for each iteration is a self-contained, auto-saved HTML file (e.g., \texttt{qual/coding/rulefile\_coding\_it\{x\}.html}), with the labels for each coder in \texttt{qual/coding/rulefiles\_coder{1,2}\_it\{x\}.csv} and the disagreements resolved in \texttt{rulefiles\_resolved\_it\{x\}.csv} (reference pattern). The script \texttt{src/qual/score\_rulefiles.py} produces the results figures that will be explained in Section~\ref{sec:res-pi2}.

\subsubsection{Qualitative data} The content of the rule files from a random sample of \textbf{150 repositories} is coded, distributed 60/30/30/30 across the four strata and internally stratified by corpus-size tercile and adoption era (before or after the AIDev cutoff). The sample size is chosen to assess saturation, not for statistical inference. These 150 repositories provide 285 individual rule files to code (some repositories have more than one), with a total corpus size of 1,866 KB (median of 5.4 KB per file); ; script \texttt{src/mine/select\_qual\_sample.py}, manifest in \texttt{data/qual\_sample\_report.json}.

The sampling unit is the repository, not the file, and the entire corpus of rule files from each selected repository is coded. Sampling individual files would reproduce in the qualitative phase the problem that deduplication solves in the quantitative phase: a repository with sixteen \texttt{.cursor/rules/*.mdc} files would contribute sixteen observations even though it is a single project. This choice also maintains the same unit of analysis in both halves of the study.

The distribution across strata is approximately equal rather than proportional, as the comparison of content \textit{across} visibility bands is one of the research questions; a proportional distribution would leave the least visible stratum with only about ten cases. This distribution comfortably meets the minimum quota outside stratum P required by the design: without it, the sample would be filled with files from popular repositories, where adoption is predictably more widespread, and conclusions about content would be limited to highly visible projects. The quota was a real risk from the outset—--it could happen that the less visible strata would not yield enough material--—and the design anticipated that, if this occurred, it \textit{would in itself be a finding regarding the gradient} and would be reported as such rather than filling the quota from stratum P. The material available after data mining (1,150 rule files outside stratum P) makes the quota achievable with room to spare.

\subsubsection{Codebook}

The codebook is deductive, derived from the framework’s mechanisms and the authors’ experience in the pilot study. It is frozen before coding begins and distributed among the coders. It contains definitions, examples, and counterexamples, as well as a set of questions that guide the coding process: \textit{Which harness mechanism(s) are materialized in this content? What type of content is it? What kind of instructions are given? What kind of rules are written? What kind of knowledge is documented? What kind of decisions are recorded?} Additionally, any content that does not fit within the framework is coded, and the reason why it does not fit is documented. Coding is open for that content, and what does not fit is recorded in the coders' notes; the one inductive category that proved reliably codeable, \texttt{proc\_substantial}, is reported in Section~\ref{sec:res-pi2}

Finally, it should be noted that the confusion between procedural and normative content is explicitly coded: instructions such as  ``how to do X'' stored in files that are always loaded, which the framework distinguishes as \textit{skills}—--procedural and on-demand--—from system specifications--—normative and always loaded---. This is a claim that currently rests on conceptual grounds; if it proves to be prevalent, it will receive direct empirical support.

\subsubsection{Categories (M1--M8)}\label{sec:metodo-categorias}

The deductive categories in the codebook replicate, one by one, the mechanisms of the harness Table~\ref{tab:mecanismos}, with the same exclusion as in RQ1: M4 (N-version mindset and parallel agents) is discarded along with H4, because it leaves no trace in the content of a rule file. A category is marked as present if \textit{any} file in the repository corpus contains it; presence is determined based on the content, never on the file name.

\begin{description}
  \item[M1 --- guidance for the agent  (H1).] The text provides guidance to the agent on the project: folder structure and where to find each item, how it is built and executed, naming and structural conventions.
  \item[M2 --- persistent knowledge  (H2).] The text records a decision and the \textit{reasoning behind it}, or refers to where it is documented; it explains why the project is the way it is.
  \item[M3 --- executable specifications  (H3).] The text sets forth an acceptance criterion, a behavioral or API contract, as a requirement that any change must satisfy; merely mentioning that specifications exist or how to execute them is not enough—that is M1.
  \item[M5 --- normative specifications  (H5).] The text states restrictions that apply to any task: modal obligations and prohibitions (\texttt{MUST}, \texttt{NEVER}).
  \item[M6 --- prescribed structured consultation (H6).] The text instructs the agent \textit{when to pause and ask} a human rather than decide on its own. Unlike H6 in  Section~\ref{sec:amenazas}, which measures whether such a consultation was actually \textit{observed} in a PR comment, M6 measures whether it is \textit{prescribed} as a written policy: these are two constructions of the same mechanism across different corpora  (Section~\ref{sec:res-pi2}), and this human coding of M6 does not depend on the lexical classifier for H6 nor is it affected by its exclusion from the quantitative study.
  \item[M7 --- acceptance based on evidence  (H7).] The text requires that concrete evidence accompany a change: tests, checklists, CI output, screenshots.
  \item[M8 --- Graduated autonomy  (H8).] The text distinguishes between what the agent can do without supervision and what requires human approval; it limits autonomy by area, risk, or scope.
\end{description}

In addition, \texttt{proc\_substantial} is coded, referring to the inductive category described in the previous subsection, which documents the confusion between procedural and normative content.

\subsubsection{Descriptive measure}\label{sec:metodo-medida-pi2}

RQ2 reports a single descriptive measure: \textit{prevalence by category}, that is, the proportion of the qualitative sample  (conditioned on H1\,=\,1 because we are working with rule files) whose content a coder labeled under each category M1–M8 or \texttt{proc\_substantial} (Section~\ref{sec:res-pi2}, Table~\ref{tab:pi2}). This is the same concept of prevalence as in RQ1 (Section~\ref{sec:metodo-agregacion}) applied to a different instrument: here, the presence indicator is assigned by a human coder after reading the entire content, not by an automatic detector based on the path or lexical pattern. For these categories, neither breadth, co-occurrence, nor temporal evolution are calculated: the qualitative sample does not aim to produce a composite index but rather to analyze the content of the rule file.


\paragraph{H6 (Structured consultation)} The lexical classifier for H6 (Section~\ref{sec:metodo-mecanismos}) is validated against human coding. Unlike categories M1--M8, which a coder assigns by reading the entire contents of a rule file, the unit here is a single PR or issue comment; the coders, the overlap and the adjudication protocol are the same as above. Hence, two independent coders label the sample defined in Section~\ref{sec:metodo-mecanismos} using a self-contained, auto-saved HTML tool (\texttt{qual/coding/consultation\_coding\_full\_coder1.html}), and a second coder repeats the first 20\,\% of the items in the same order (\texttt{consultation\_coding\_full\_coder2.html}) for the reliability calculation.
The resulting labels (\texttt{qual/coding/labels\_coder1\_full.csv}, \texttt{labels\_coder2\_full.csv}) are scored using  \texttt{src/classify/score\_validation\_full.py}, which produces  \texttt{data/c6\_validation\_full.json} with the prevalence, \textit{recall}, and final decision. The decision rule regarding this result is also established before coding: if there are fewer than ten confirmed positives, \textit{recall} cannot be meaningfully estimated and is not calculated; instead, prevalence is reported along with its interval, and it is explicitly stated that the classifier’s \textit{recall} \textit{cannot be estimated at that prevalence}
. With ten or more positives, \textit{recall} is calculated, but if it falls below 0.70, H6 is excluded from the study rather than replacing the method. The inter-coder agreement and the result obtained are reported in Section~\ref{sec:res-h6-validacion}, and the qualitative interpretation of that result is discussed in Section~\ref{sec:amenazas}.

\subsection{Reproducibility and Content Handling}\label{sec:metodo-repro}

The replication package includes the instrument (see Section~\ref{sec:data-availability}). We publish the code the article cites, the codebook, every coding label together with the third author's adjudications, the human-validation sample of the consultation detector, and content-addressed pointers (repository, path, commit, blob SHA) to every coded rule file, so that any of them can be retrieved from its source and verified by hash. The sampling frame is not shipped: it regenerates deterministically from the pinned dataset and the seed recorded in the script. Each figure and table in the article indicates the script used to generate it.

No third-party content is republished beyond brief citations covered by fair use: the mined repositories have heterogeneous licenses, and a considerable fraction does not declare any. The text of the rule files is analyzed locally. Verbatim citations appearing in the article will preferably come from repositories with a declared license, and the license composition of the qualitative sample is reported.

The unit of reporting is the repository. Results referring to identifiable individuals are not published, nor are profiles of individual developers constructed, even though the source data would allow for it.

\section{Results}\label{sec:resultados}

\subsection{Sampling design}\label{sec:res-muestra}

Of the 5,807 repositories in the framework, \textbf{5,435} (93.6\,\%) yielded a usable clone. Of the remaining 372, the API confirmed that 39 repositories still existed—renamed, archived, or accessible under the same name—but neither the initial clone nor, where applicable, the retry following name resolution were completed, and thus they are not included in the analyzed corpus; the remaining 333 were inaccessible due to deletion or being set to private. All repositories in the framework end up in one of these three states, and the complete record is published (see Section~\ref{sec:data-availability}).

Attrition increases monotonically as we move down through the visibility bands: 2.7\,\% in P, 3.5\,\% in G1, 11.8\,\% in G2, and 14.4\,\% in G3  (Table~\ref{tab:disposicion}). The interpretation of this progression is discussed in Section~\ref{sec:amenazas}, because it determines how the lowest stratum (G3) should be interpreted. Across the cloned repositories, 35,749 artifacts were identified that implement these mechanisms (rule or context files, ADRs, executable specifications, PR templates, or CODEOWNERS) after deduplication. This figure is the output of the corpus mining (the file tree and \texttt{git} history of the 5,435 repositories, Section~\ref{sec:metodo-dedup}), not a measure of the size of that corpus. A total of 310 symbolic links, 20 artifacts with indeterminable dates, and 737 byte-for-byte duplicates were discarded (1,067 in total; script \texttt{src/mine/mine\_harness.py}; manifest in \texttt{data/prevalence\_report.json}, field \texttt{artifacts\_dropped\_A2}); the raw count (before deduplication) broken down by stratum is in Appendix~\ref{sec:apx-aidev}, Table~\ref{tab:apx-aidev}. Another 240 stub-files are marked, but they are not subtracted from that total: they only affect the prevalence of H1, whose sensitivity to that marker is measured separately (Section~\ref{sec:res-sensibilidad}). The magnitude of exact duplicates --- 2\,\% of raw artifacts—--- confirms that deduplication is necessary. Table~\ref{tab:artefactos-mecanismo} breaks down the total number of artifacts by mechanism, both raw and after deduplication, and shows that H3 dominates the total: a single repository with a Gherkin test suite can contain hundreds of *.feature files, whereas prevalence (Table~\ref{tab:prevalencia}) counts repositories, not files. Rule files (H1) are, in fact, a minority of the total, and  Table~\ref{tab:prevalencia} shows that their prevalence is greater than that of any other mechanism. Section~\ref{sec:res-prevalencia} analyzes prevalence by stratum and by mechanism.

\begin{table}[h]
\caption{Sample distribution by stratum: from each repository in the framework to its terminal state. ``API-Fallback'' refers to repositories whose cloning failed and whose status was resolved by querying the GitHub API (renamed, archived, or accessible under the same name) without obtaining a usable clone ---a renamed repository whose cloning retry succeeds is counted as \textit{Cloned}, not here---; ``Unavailable'' refers to repositories that were deleted or made private. Attrition $=$ (API-Fallback $+$ Unavailable) / Framework.}\label{tab:disposicion}
\footnotesize
\begin{tabular*}{\textwidth}{@{\extracolsep\fill}lrrrrr@{\extracolsep\fill}}
\toprule
Stratum & Frame & Cloned & API Fallback & Unavailable & Attrition \\
\midrule
P  & 2,807 & 2,732 & 34 & 41  & 2.7\,\% \\
G1 & 1,000 & 965   & 4  & 31  & 3.5\,\% \\
G2 & 1,000 & 882   & 1  & 117 & 11.8\,\% \\
G3 & 1,000 & 856   & 0  & 144 & 14.4\,\% \\
\midrule
\textbf{Total} & \textbf{5,807} & \textbf{5,435} & \textbf{39} & \textbf{333} & \textbf{6.4\,\%} \\
\botrule
\end{tabular*}
\footnotesize\emph{Source: src/mine/mine\_harness.py \texttt{->} data/harness\_\{P,G1,G2,G3\}\_raw.json, aggregated by  src/describe/prevalence.py -> data/prevalence\_report.json (attrition\_by\_stratum).}
\end{table}

\begin{table}[h]
\caption{Artifacts per mechanism, raw and after deduplication (symlinks, exact duplicates, and those without a resolvable date are discarded; H1 stub-files are marked but not subtracted from this total—they are excluded separately when calculating prevalence, Section~\ref{sec:metodo-agregacion}). H3 dominates the total: a single repository with a Gherkin test suite can contribute hundreds of \texttt{*.feature}, whereas prevalence 
(Table~\ref{tab:prevalencia}) counts repositories, not files --- rule files (H1) are, in fact, a minority of the total.}\label{tab:artefactos-mecanismo}
\footnotesize
\begin{tabular*}{\textwidth}{@{\extracolsep\fill}lrr@{\extracolsep\fill}}
\toprule
Mechanism & Raw & After deduplication \\
\midrule
H3 Executable specifications & 29,675 & 28,946 \\
H1 Context engineering & 4,146 & 3,810 \\
H7 Evidence-based acceptance & 1,164 & 1,164 \\
H2 Persistent shared knowledge & 1,030 & 1,029 \\
H8 Graduated autonomy & 801 & 800 \\
\midrule
\textbf{Total} & \textbf{36,816} & \textbf{35,749} \\
\botrule
\end{tabular*}
\footnotesize\emph{Source: src/describe/mechanism\_breakdown.py \texttt{->} data/mechanism\_breakdown.json (reuses \texttt{build\_matrix.dedupe()}, the same deduplication rule that produces data/prevalence\_report.json).}
\end{table}

\subsection{RQ1: Observable adoption}\label{sec:res-rq1}

This section addresses RQ1 across five dimensions of analysis: prevalence (including the normative content H5, conditional on H1), visibility gradient, breadth, co-occurrence, and temporal adoption (which combines the coverage of the observed work and when it was adopted). We also report the validation against human coding of H6, which was dropped from the quantitative analysis (Section~\ref{sec:res-h6-validacion}). Finally, the sensitivity of these dimensions is described.

\subsubsection{Prevalence}\label{sec:res-prevalencia}

Table~\ref{tab:prevalencia} presents the main results of the quantitative study: the prevalence of each mechanism by stratum, with 95\,\% Wilson intervals. The reweighted row estimates the population of 116,211 repositories with agentic activity. When reweighted to the 116,211 repositories, the adoption rate of \textit{any} mechanism is 21.7\,\%. The 65.3\,\% of the \textit{popular} stratum represents the view from above: it describes the 2,807 most visible repositories, which account for 2.4\,\% of the population  (Table~\ref{tab:estratos}). Any claim about ``the adoption of the harness'' that relies on popular repositories is, by construction, looking at the wrong tail of the distribution.

\begin{table}[h]
\caption{Prevalence of each mechanism by stratum (estimating E1: the artifact exists in the current \texttt{HEAD}), with 95\,\% Wilson confidence intervals. Values are expressed as percentages (\%), with their corresponding confidence intervals reported below in brackets [...]. Finally, the prevalence of H5  (normative content) is reported below as an indicator conditioned on H1
}\label{tab:prevalencia}
\footnotesize
\begin{tabular*}{\textwidth}{@{\extracolsep\fill}lrrrrrrr@{\extracolsep\fill}}
\toprule
Stratum & $n$ & H1 & H2 & H3 & H7 & H8 & Any \\
\midrule
P  & 2,732 & 51.2 & 1.9 & 6.8 & 32.5 & 22.4 & 65.3 \\
   &       & \tiny[49.3--53.1] & \tiny[1.5--2.5] & \tiny[5.9--7.8] & \tiny[30.7--34.2] & \tiny[20.9--24.0] & \\
G1 & 965   & 36.3 & 0.9 & 3.8 & 12.2 & 11.6 & 46.4 \\
   &       & \tiny[33.3--39.4] & \tiny[0.5--1.8] & \tiny[2.8--5.2] & \tiny[10.3--14.4] & \tiny[9.7--13.8] & \\
G2 & 882   & 26.8 & 1.0 & 2.5 & 7.9  & 5.6  & 33.7 \\
   &       & \tiny[23.9--29.8] & \tiny[0.5--1.9] & \tiny[1.6--3.8] & \tiny[6.3--9.9] & \tiny[4.2--7.3] & \\
G3 & 856   & 12.6 & 0.6 & 1.3 & 5.3  & 2.6  & 17.4 \\
   &       & \tiny[10.6--15.0] & \tiny[0.2--1.4] & \tiny[0.7--2.3] & \tiny[4.0--7.0] & \tiny[1.7--3.9] & \\
\midrule
\textbf{P/G3} & & \textbf{4.1} & 3.3\textsuperscript{a} & \textbf{5.3} & \textbf{6.2} & \textbf{8.7} & \textbf{3.7} \\
\midrule
\textbf{Reweighted} & \textit{116,211} & \textbf{16.2} & \textbf{0.7} &
  \textbf{1.7} & \textbf{6.5} & \textbf{3.7} & \textbf{21.7} \\
\botrule
\end{tabular*}
\footnotesize\emph{Source: src/describe/prevalence.py \texttt{->} data/prevalence\_report.json.
The P/G3 row is computed on the full percentages in that file, not on the rounded figures in this table, so dividing
the published columns may give a different last digit.
\textsuperscript{a}H2's Wilson intervals overlap across G1, G2, and
G3 (Section~\ref{sec:amenazas}); the ratio is reported for completeness but is not a statistically resolved difference.}
\end{table}

\paragraph{Normative content (H5, conditioned on H1)}\label{sec:res-normativo} Among repositories with rule files, the percentage whose files contain at least one normative clause is \textbf{90.9}\,\% in P (1,271/1,399), 85.4\,\% in G1 (299/350), 82.2\,\% in G2 (194/236), and 74.1\,\% in G3 (80/108) (see Table~\ref{tab:h5-idioma}). Two interpretations. The first: \textit{writing rules is what one does with a rules file}. This practice is so widespread among those who adopt H1 that the indicator behaves as a constant, and for that reason it cannot be used to distinguish repositories; that is, it is reported as a description, not as a variable. The second: \textit{even so, it decreases monotonically with visibility}.

Much of that decline is due to a flaw in the tool. The H5 detector identifies modal clauses based solely on English patterns (\texttt{MUST}, \texttt{NEVER}, do not. . . ). A rule file written in Japanese that states its requirements in Japanese does not match anything and is recorded as lacking normative content. Automatic language detection (Section~\ref{sec:metodo-idioma}) allows this detection error to be narrowed down rather than simply reported..

Of the 248 repositories whose H1 artifacts all scored zero clauses—--the set where the error can occur—--the fraction that the instrument cannot read, either due to being in another language or remaining indeterminate, is 14.1\,\% in P, 28.0\,\% in G1, 16.7\,\% in G2, and 39.3\,\% in G3 (Table~\ref{tab:h5-idioma}, column $n$). Out of the 276 files contributed by these 248 repositories, 215 are in English (genuine zeros: they contain no modal clauses, even though the detector could read them) and 61 represent the failure of the instrument itself, as detailed in Table~\ref{tab:idiomas}. This is not a single-language phenomenon.

\begin{table}[h]
\caption{Fraction of repositories with rule files that contain at least one normative clause, before and after excluding repositories whose files the instrument cannot read from the denominator. The correction is greatest precisely where the gradient was most pronounced. $n$ is the total number of repositories with H1 artifacts with zero clauses in that stratum (denominator of the Excluded column).}\label{tab:h5-idioma}
\footnotesize
\begin{tabular*}{\textwidth}{@{\extracolsep\fill}lrrrrr@{\extracolsep\fill}}
\toprule
Stratum & $n$ & Published & Excluded & Corrected & Shift \\
\midrule
P  & 128 & 90.9\,\% & 18 & 92.0\,\% & $+1.2$ \\
G1 & 50  & 85.4\,\% & 14 & 89.0\,\% & $+3.6$ \\
G2 & 42  & 82.2\,\% & 7  & 84.7\,\% & $+2.5$ \\
G3 & 28  & 74.1\,\% & 11 & 82.5\,\% & $\mathbf{+8.4}$ \\
\midrule
\multicolumn{1}{l}{\emph{Breadth}} & & 16.8 pts & & 9.5 pts & $-43\,\%$ \\
\botrule
\end{tabular*}
\footnotesize\emph{Source: src/qual/h5\_language\_sensitivity.py \texttt{->} data/h5\_language\_sensitivity.json.}
\end{table}

\begin{table}[h]
\caption{Dominant language of the 276 H1 files across the 248 repositories with zero clauses. ``No ingles'' groups together languages with fewer than two files.}\label{tab:idiomas}
\footnotesize
\begin{tabular*}{\textwidth}{@{\extracolsep\fill}lrr@{\extracolsep\fill}}
\toprule
Language & Files & \% \\
\midrule
English (genuine zero) & 215 & 77.9\,\% \\
japanese & 22 & 8.0\,\% \\
Chinese (zh-cn) & 15 & 5.4\,\% \\
Indeterminate & 5 & 1.8\,\% \\
Korean & 5 & 1.8\,\% \\
Russian & 3 & 1.1\,\% \\
Portuguese & 3 & 1.1\,\% \\
French & 2 & 0.7\,\% \\
Italian & 2 & 0.7\,\% \\
Vietnamese & 1 & 0.4\,\% \\
Norwegian & 1 & 0.4\,\% \\
Estonian & 1 & 0.4\,\% \\
German & 1 & 0.4\,\% \\
\midrule
Total & 276 & 100.0\,\% \\
\botrule
\end{tabular*}
\footnotesize\emph{Source: src/qual/h5\_language\_sensitivity.py \texttt{->} data/h5\_language\_sensitivity.json (zero\_clause\_repos.per\_file\_language).}
\end{table}

The gradient breadth drops from 16.8 points to 9.5 (Table~\ref{tab:h5-idioma}, columns \emph{Published} and \emph{Corrected}, respectively). To put it bluntly: about two-fifths of the apparent decrease in regulatory content as we move down the visibility bands was not because the drafts contained fewer regulations, but because the tool was unable to read them. What remains after the correction is still a steady decline, and that part is indeed a finding: the artifacts from the less visible repositories are also \textit{lighter} in terms of regulatory content. But the unadjusted published magnitude overestimates the effect by nearly double, and the two figures are reported together for that reason.

\subsubsection{Visibility gradient}\label{sec:res-forma}

The gradient exists and is monotonic across all five mechanisms. Without a single exception, each mechanism is more prevalent as the band becomes more visible  (Table~\ref{tab:prevalencia}). For H1, H3, H7, and H8, the intervals of adjacent strata do not overlap, so the decline is not attributable to sampling. H2 is the exception, and not in the direction one might expect. Its intervals overlap completely between G1 [0.5–1.8], G2 [0.5–1.9], and G3 [0.2–1.4]: \emph{there is no discernible gradient} gradient because there is virtually nothing to graduate. Persistent knowledge (decision logs, project memory) is practically absent across the entire population, including the most visible segment, where it reaches  1.9\,\%. The P/G3 ratio of  $\times$3.3 should not be interpreted as a gradient (Table~\ref{tab:prevalencia}).

The shape of the gradient is interpretable. The mechanisms do not decline at the same rate. From the steepest to the gentlest decline between P and G3 (row P/G3, Table~\ref{tab:prevalencia}): \textbf{H8} \texttt{CODEOWNERS} ($\times$8.7), \textbf{H7} PR template  ($\times$6.2), \textbf{H3} executable specifications ($\times$5.3), \textbf{H1} rule files ($\times$4.1) and \textbf{H2} ($\times$3.3, not resolvable). The two that drop the most are precisely the two that presuppose a team: \texttt{CODEOWNERS} assigns reviewers by path, and a PR template structures communication between the contributor and the reviewer. Neither makes sense for a single person. The mechanism that holds up best is rule files, which an individual developer writes exactly the same way as a team would: it is an artifact directed at the agent, not at other humans. It is therefore possible that the visibility gradient is measuring, to a substantial extent, the \textit{presence of a team rather than the adoption of a harness}. This is an interpretation and is offered as such: verifying it requires a direct measure of team size that this study does not include, and it is noted as future work. What remains an empirical fact is that harness mechanisms designed for individual use and those requiring human coordination behave differently as visibility decreases, and the conceptual framework did not anticipate that distinction.

\subsubsection{Breadth}\label{sec:res-amplitud}

The breadth is defined as the count of mechanisms present among the five common ones (H1, H2, H3, H7, and H8). Table~\ref{tab:amplitud} breaks down the repositories by the number of mechanisms adopted. The pattern is unmistakable: the adoption of the harness as a coherent whole is very low across all visibility bands. Even in the most visible stratum, only 2.6\,\% of repositories incorporate four or five mechanisms; outside of it, the figure drops to three repositories in G1, 5 in G2, and \textit{none} in G3. The mode is zero mechanisms in three of the four strata, and one in the remaining stratum.

\begin{table}[h]
\caption{Breadth distribution (number of mechanisms out of the five common ones). Row percentages. ``Mean'' refers to the mean breadth of the stratum.}\label{tab:amplitud}
\footnotesize
\begin{tabular*}{\textwidth}{@{\extracolsep\fill}lrrrrrrrr@{\extracolsep\fill}}
\toprule
Stratum & $n$ & 0 & 1 & 2 & 3 & 4 & 5 & Mean \\
\midrule
P  & 2,732 & 34.7\,\% & 32.1\,\% & 19.7\,\% & 10.9\,\% & 2.2\,\% & 0.4\,\% & 1.15 \\
G1 & 965   & 53.6\,\% & 31.7\,\% & 11.3\,\% & 3.1\,\%  & 0.3\,\% & ---     & 0.65 \\
G2 & 882   & 66.3\,\% & 26.5\,\% & 5.0\,\%  & 1.6\,\%  & 0.3\,\% & 0.2\,\% & 0.44 \\
G3 & 856   & 82.6\,\% & 13.6\,\% & 2.8\,\%  & 1.1\,\%  & ---     & ---     & 0.22 \\
\midrule
\textbf{Reweighted} & \textit{116,211} & & & & & & & \textbf{0.29} \\
\botrule
\end{tabular*}
\footnotesize\emph{Source: src/describe/prevalence.py \texttt{->} data/prevalence\_report.json (breadth\_distribution, mean\_breadth).}
\end{table}

\subsubsection{Co-occurrence}\label{sec:res-coocurrencia}

\begin{figure}[h]
\centering
\includegraphics[width=0.55\textwidth]{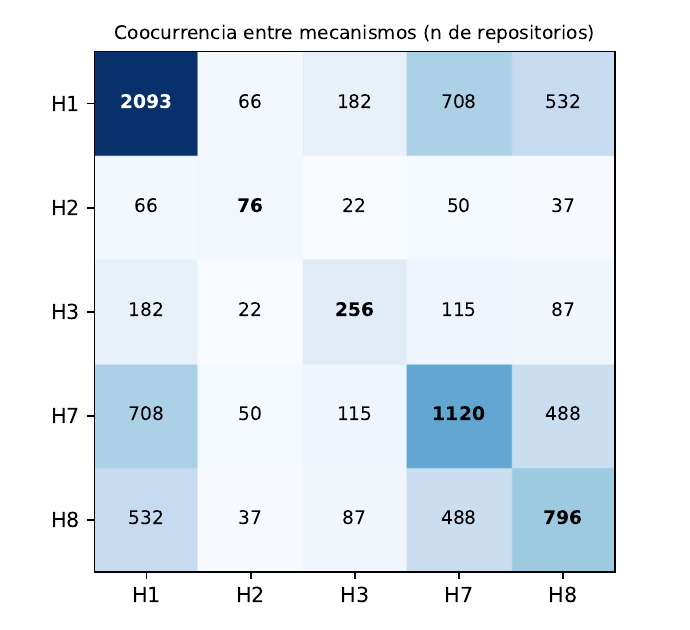}
\caption{Co-occurrence matrix among the five common mechanisms (raw counts out of the 5,435 cloned repositories, unreweighted). The diagonal represents the raw prevalence of each individual mechanism}\label{fig:cooccurrence}
\footnotesize\emph{Source: src/describe/plots.py}
\end{figure}

The co-occurrence matrix shows that the most frequent combinations are  H1\,\&\,H7 (708 repositories), H1\,\&\,H8 (532) and H7\,\&\,H8 (488): that is, rule files accompanied by governance artifacts that many projects already possessed. Combinations involving strictly methodological mechanisms are rare: H1\,\&\,H3 appears in  182 repositories, and H1\,\&\,H2 in 66, out of 5,435. Figure~\ref{fig:cooccurrence} shows the complete matrix of the ten possible pairs among the five common mechanisms, with the diagonal indicating the raw prevalence of each mechanism separately; H5 is excluded from this matrix because it is not an independent mechanism (Section~\ref{sec:metodo-h5}).

These results point to \textit{limited and uneven adoption}, while providing greater precision: what is predominantly adopted is an \textit{artifact}—--rule files--—rather than a \textit{method}. The mechanisms that require sustained methodological investment (persistent knowledge, executable specifications) remain marginal even where rule files are already in place.

\subsubsection{Temporal adoption}\label{sec:res-adopcion-temporal}

\paragraph{Coverage of observed work (E2)}\label{sec:res-e2} The previous results answer E1: how many repositories possess each mechanism at the time of cloning. E2 answers the other question—what fraction of the recorded agentic work (the PR) occurred under a harness—by evaluating each mechanism at the moment each PR was created. It is calculated based on the 33,596 agentic PRs in stratum P (December 2024 to July 2025); AIDev does not publish PR tables outside the \textit{popular} tier, so E2 is not estimated for the G strata, which is explicitly declared as a limitation of the scope. Table~\ref{tab:e2} compares E2 and E1 by mechanism; the E1 column in that table does not match the prevalence in Table~\ref{tab:prevalencia} because one is weighted by PR and the other by repository, not because they measure different things  (Appendix~\ref{sec:apx-e2} breaks this down with an example).

\begin{table}[h]
\caption{E2 versus E1 across the 33,596 agentic PRs in stratum P. Both columns are weighted by PR, meaning that E1 here (55.7\,\% for H1) does not match the prevalence by repository in  
Table~\ref{tab:prevalencia} (51.2\,\%): repositories with more agentic activity carry more weight.}\label{tab:e2}
\footnotesize
\begin{tabular*}{\textwidth}{@{\extracolsep\fill}lrrr@{\extracolsep\fill}}
\toprule
Mechanism & E2 (at the moment of the PR) & E1 (present at cloning) & Ratio \\
\midrule
H1 & 26.4\,\% & 55.7\,\% & 2.11 \\
H2 & 0.7\,\%  & 1.4\,\%  & 2.12 \\
H3 & 3.1\,\%  & 6.3\,\%  & 2.00 \\
H7 & 28.9\,\% & 34.1\,\% & 1.18 \\
H8 & 15.2\,\% & 18.0\,\% & 1.18 \\
\midrule
Any & 43.5\,\% & 61.2\,\% & 1.41 \\
Difference E1(Any) $-$ E2(Any) & \multicolumn{2}{c}{\textbf{17.7\,\% (5,954 PRs)}} & \\
Mean breadth & 0.74 & 1.15 & \\
\botrule
\end{tabular*}
\footnotesize\emph{Source: src/describe/coverage\_e2.py \texttt{->} data/coverage\_e2\_report.json. The
``Difference'' row es E1(Any) $-$ E2(Any) $=$ 61.2 $-$ 43.5, expressed in PRs and percentage points; this is the figure cited in the text as PRs counted as harnessed at cloning when they were not harnessed when opened}
\end{table}

A total of 17.7\,\% of the agentic PRs (5,954) would be counted as  ``harnessed'' by looking at the current state of the repository, but they were not when they were opened. This is the magnitude of the error committed by a design that evaluates the presence of the artifact at the time of data collection, and the reason why a study with outcome variables must evaluate at the exact moment of the PR.

The most informative detail lies in the ratio column. The mechanisms specific to the agentic era (rule files, persistent knowledge, executable specifications) approximately doubled between the observed work and cloning  ($\times$2.00 a $\times$2.12). The two pre-existing governance artifacts barely moved ($\times$1.18): they were already there. The harness did not grow uniformly; it grew in the area that is specifically a response to the introduction of agents.

\paragraph{Adoption timing.}\label{sec:res-adopcion} Of the 3,810 H1 artifacts with a determinable introduction date  (Table~\ref{tab:artefactos-mecanismo}), 58.3\,\% appeared after July 30, 2025 (script \texttt{src/describe/prevalence.py}; manifest in \texttt{data/prevalence\_report.json}, field \texttt{hyp\_c\_h1\_after\_aidev\_cutoff}), that is, after the closure of the PRs window on which the sampling frame was defined. The practice measured by this study is, for the most part, subsequent to the agentic work that gave rise to the sample.

\begin{figure}[h]
\centering
\includegraphics[width=0.85\textwidth]{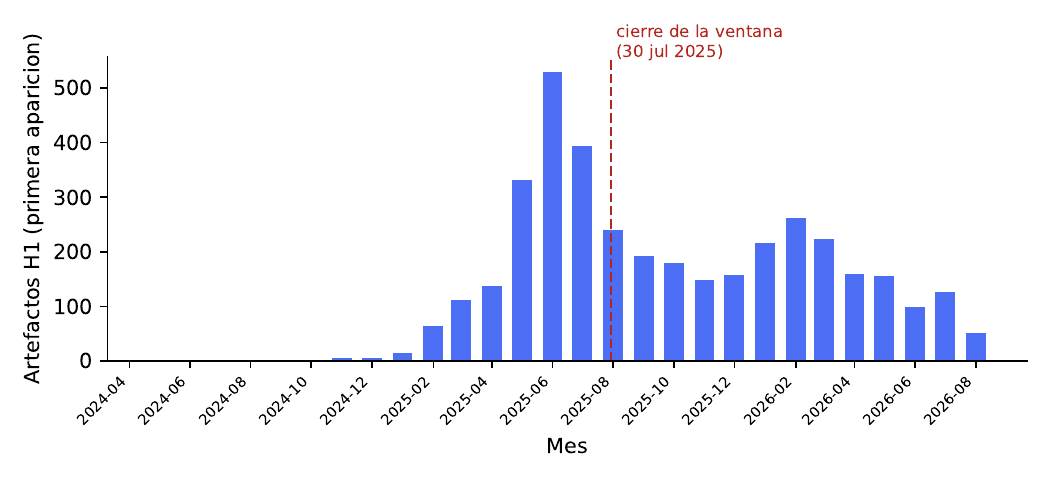}
\caption{Monthly first appearances of H1 (rule files), with the agentic PR window closure (July 30, 2025) marked in red. The shape of the curve should not be interpreted directly: two effects distort it (saturation and selection), which are discussed in the text.}\label{fig:adopcion}
\footnotesize\emph{Source: src/describe/plots.py}
\end{figure}

The monthly histogram of first appearances rises from early 2025, peaks in June 2025 (530 artifacts), declines, and rises again in January–March 2026 (217, 261, and 224). Figure~\ref{fig:adopcion} shows the complete histogram, with the end of the sampling window marked.

The shape of this curve should not be interpreted directly, and in particular, its peak does not indicate that adoption has reached a ceiling. Two effects distort it. The first is saturation: \textit{first} appearances are counted, and a repository can only appear once, so the raw count necessarily declines as the set of repositories that have not yet adopted is exhausted, even if the propensity to adopt remains constant. The second is selection bias: the framework requires agentic activity prior to August 2025, and it is plausible that adopting a rule file coincides with beginning to work with agents, which artificially concentrates first appearances just before the window closes. A correct interpretation requires cumulative incidence among the at-risk population, with right-censored data at the cloning date, which is noted as a future work.

What is open to interpretation, however—--since it lies outside the scope of both effects—--is the uptick in early 2026, six months after the window closed: adoption did not stop at the end of the period defined by the sample.

\subsubsection{Validation of H6 (structured consultation)}\label{sec:res-h6-validacion}

Two independent coders labeled 536 PR and issue comments (Section~\ref{sec:metodo-medida-pi2}) and agreed on the 107-item overlap at $\kappa = 0{.}681$ (substantial). Measured against those labels, its lexical detector recovered 26 of the 53 consultations the coders confirmed --- recall 0.491 and precision 0.366 over its 71 positives, the strict and lax blocks of Table~\ref{tab:h6-cajones} --- below the 0.70 recall threshold fixed in advance. Under that rule H6 is excluded from the quantitative analysis rather than re-detected by another method (that is why H6 does not appear in Table~\ref{tab:prevalencia}). The mechanism itself is real: 53 structured consultations were confirmed in the sample (Table~\ref{tab:h6-cajones}). What the detector missed, and why, is discussed in Section~\ref{sec:amenazas}.

\begin{table}[h]
\caption{Sampling design and results of the human validation of H6, by classifier block. ``Confirmed'' refers to the number of actual structured consultations according to the lead coder.}\label{tab:h6-cajones}
\footnotesize
\begin{tabular*}{\textwidth}{@{\extracolsep\fill}lrrlr@{\extracolsep\fill}}
\toprule
Block & Population & Coded & Design & Confirmed \\
\midrule
Strict detection            & 8      & 8   & census            & 2  \\
Lax detection                & 63     & 63  & census            & 24 \\
Question, without options       & 165    & 165 & census            & 27 \\
Options, without question        & 6,414  & 150 & sample ($n{=}150$) & 0  \\
No signal                 & 49,898 & 150 & sample ($n{=}150$) & 0  \\
\midrule
Total                         & 56,548 & 536 &                  & 53 \\
\botrule
\end{tabular*}
\footnotesize\emph{Source: src/classify/consultation\_full.py \texttt{->}
data/consultation\_report\_full.json, data/consultation\_validation\_sample\_full.csv.}
\end{table}

\subsubsection{Sensitivity Analysis}\label{sec:res-sensibilidad}

\paragraph{Stub-files} Counting them as H1 present instead of excluding them from prevalence alters the ''any mechanism'' figure by less than half a percentage point across all four strata (65.4\,\% versus 65.3\,\% in P; 17.6\,\% versus 17.4\,\% in G3), as fully shown in Table~\ref{tab:sens-stubs}. Of the four data cleaning criteria from Section~\ref{sec:metodo-dedup} (symbolic links, unresolvable dates, exact duplicates, and stub-files), this is the only one that involves a debatable judgment--—the other three being mechanically verifiable—--and it turns out to be irrelevant to the conclusions.

\begin{table}[h]
\caption{Sensitivity to the stub-file filter: published prevalence of "any mechanism" (excluding stub-files) versus retaining them, by stratum}\label{tab:sens-stubs}
\footnotesize
\begin{tabular*}{\textwidth}{@{\extracolsep\fill}lrrr@{\extracolsep\fill}}
\toprule
Stratum & Published & With stub-files & Shift \\
\midrule
P  & 65.3\,\% & 65.4\,\% & $+0.1$ \\
G1 & 46.4\,\% & 46.8\,\% & $+0.4$ \\
G2 & 33.7\,\% & 34.1\,\% & $+0.4$ \\
G3 & 17.4\,\% & 17.6\,\% & $+0.2$ \\
\midrule
\textbf{Reweighted} & \textbf{21.7\,\%} & \textbf{22.0\,\%} & \textbf{$+0.3$} \\
\botrule
\end{tabular*}
\footnotesize\emph{Source: src/describe/prevalence.py \texttt{->} data/prevalence\_report.json (sensitivity\_stubs\_kept).}
\end{table}

\paragraph{Projects with a Single Agentic PR} When restricted to repositories with at least two agentic PRs, all figures increase  (73.4\,\% in P, 50.0\,\% in G1, 38.5\,\% in G2, 23.8\,\% in G3, and 27.9\,\% in the reweighted estimate), but the order and shape of the gradient remain unchanged (Table~\ref{tab:sens-ge2}). The long tail of single-use projects dilutes the level of all estimates; it does not create the gradient. This was the sensitivity check with the potential to overturn the main result, and it fails to do so.

\begin{table}[h]
\caption{Sensitivity to the single-agentic-PR project filter: published prevalence of "any mechanism" (all repositories) versus restricting to repositories with at least two agentic PRs, by stratum.}\label{tab:sens-ge2}
\footnotesize
\begin{tabular*}{\textwidth}{@{\extracolsep\fill}lrrrr@{\extracolsep\fill}}
\toprule
Stratum & $n$ ($\geq$2 PRs) & Published & $\geq$2 agentic PRs & Shift \\
\midrule
P  & 1,523 & 65.3\,\% & 73.4\,\% & $+8.1$ \\
G1 & 528   & 46.4\,\% & 50.0\,\% & $+3.6$ \\
G2 & 496   & 33.7\,\% & 38.5\,\% & $+4.8$ \\
G3 & 432   & 17.4\,\% & 23.8\,\% & $+6.4$ \\
\midrule
\textbf{Reweighted} & \textit{2,979} & \textbf{21.7\,\%} & \textbf{27.9\,\%} & \textbf{$+6.2$} \\
\botrule
\end{tabular*}
\footnotesize\emph{Source: src/describe/prevalence.py \texttt{->} data/prevalence\_report.json (sensitivity\_ge2\_agentic\_prs).}
\end{table}

\subsubsection{Answer to RQ1}\label{sec:res-respuesta-rq1}
The methodological harness is observable in practice, but its adoption is narrow, uneven, and heavily dependent on the repository's visibility. Individual mechanisms occur substantially more often than coherent configurations of multiple mechanisms, and the mechanisms specific to the agentic era remain considerably less prevalent than rule files and pre-existing governance artifacts.

\subsection{RQ2: Correspondence with the Framework}\label{sec:res-pi2}

The previous sections detect mechanisms by file path or, in the case of H5, by a lexical pattern. RQ2 manually codes the complete content of the corpus of rule files from \textbf{150 repositories with at least one rule file} (60 P / 30 G1 / 30 G2 / 30 G3, Section~\ref{sec:metodo-cualitativo}), using eight deductive categories that replicate the mechanisms of the conceptual framework—--M1 through M8, with M4 excluded just like H4—--plus \texttt{proc\_substantial} (substantial procedural content housed in a file that is always loaded, Section~\ref{sec:metodo-cualitativo}). Unlike  Table~\ref{tab:prevalencia}, which asks \textit{whether the artifact exists}, RQ2 asks  \emph{what is inside it}. 

\subsubsection{Reliability}\label{sec:res-pi2-fiabilidad} After a second iteration, all eight fields exceed the threshold $\kappa \geq 0{.}60$ on the 30-item overlap coded by both coders: M2 $\kappa=0{.}82$, M3 $\kappa=0{.}61$, M6 $\kappa=0{.}79$, M7 $\kappa=0{.}66$, M8 $\kappa=0{.}71$ and \texttt{proc\_substantial} $\kappa=0{.}68$; M1 and M5 are validated through raw agreement (93\,\% and 97\,\%) due to extreme prevalence in the overlap. 
The iteration-by-iteration breakdown, including the codebook amendment between them, is reported in Table~\ref{tab:kappa-it1} and Table~\ref{tab:kappa-it2}.

\begin{table}[h]
\caption{Inter-coder reliability for RQ2, \textbf{iteration 1} (codebook v1, 30-item overlap). ``Gate'' is the pass criterion: $\kappa$ (threshold
$0{.}60$) or raw agreement ($0{.}90$ threshold) when the baseline prevalence is extreme  ($\geq$85\,\%). PS = \texttt{proc\_substantial}.}\label{tab:kappa-it1}
\footnotesize
\begin{tabular*}{\textwidth}{@{\extracolsep\fill}lrrrll@{\extracolsep\fill}}
\toprule
Category & Base rate & $\kappa$ & Raw agreement & Gate & Result \\
\midrule
M1 & 100.0\,\% & --    & 100.0\,\% & Raw agreement & pass \\
M2 & 26.7\,\%  & 0.659 & 86.7\,\%  & $\kappa$      & pass \\
M3 & 18.3\,\%  & 0.255 & 76.7\,\%  & $\kappa$      & \textbf{fail} \\
M5 & 93.3\,\%  & 1,000 & 100.0\,\% & Raw agreement & pass \\
M6 & 11.7\,\%  & 0.526 & 90.0\,\%  & $\kappa$      & \textbf{fail} \\
M7 & 86.7\,\%  & 0.714 & 93.3\,\%  & Raw agreement & pass \\
M8 & 16.7\,\%  & 0.302 & 80.0\,\%  & $\kappa$      & \textbf{fail} \\
PS & 30.0\,\%  & 0.415 & 73.3\,\%  & $\kappa$      & \textbf{fail} \\
\botrule
\end{tabular*}
\footnotesize\emph{Source: src/qual/score\_rulefiles.py \texttt{->} data/pi2\_report\_it1.json.
Four out of eight categories fail (M3, M6, M8, PS); the codebook is amended and recoded in iteration 2.}
\end{table}

\begin{table}[h]
\caption{Inter-coder reliability for RQ2, \textbf{iteration 2} (amended codebook v2, 30-item overlap, after applying five minor transcription corrections in R031/R041/R049/R057). Same columns as 
Table~\ref{tab:kappa-it1}. \texttt{proc\_substantial} was not coded directly: it is derived from each coder's three-way \texttt{orientation} label (\texttt{MIXED}/\texttt{PROC} $\rightarrow$ 1, \texttt{NORM} $\rightarrow$ 0), so its $\kappa$ measures agreement between derived values.}\label{tab:kappa-it2}
\footnotesize
\begin{tabular*}{\textwidth}{@{\extracolsep\fill}lrrrll@{\extracolsep\fill}}
\toprule
Category & Base rate & $\kappa$ & Raw agreement & Gate & Result \\
\midrule
M1 & 93.3\,\% & 0.474 & 93.3\,\%  & Raw agreement & pass \\
M2 & 23.3\,\% & 0.815 & 93.3\,\%  & $\kappa$      & pass \\
M3 & 15.0\,\% & 0.609 & 90.0\,\%  & $\kappa$      & pass \\
M5 & 95.0\,\% & 0.651 & 96.7\,\%  & Raw agreement & pass \\
M6 & 20.0\,\% & 0.793 & 93.3\,\%  & $\kappa$      & pass \\
M7 & 55.0\,\% & 0.664 & 83.3\,\%  & $\kappa$      & pass \\
M8 & 21.7\,\% & 0.706 & 90.0\,\%  & $\kappa$      & pass \\
PS & 51.7\,\% & 0.675 & 83.3\,\%  & $\kappa$      & pass \\
\botrule
\end{tabular*}
\footnotesize\emph{Source: src/qual/score\_rulefiles.py \texttt{->} data/pi2\_report\_it2.json.
All eight categories exceed the threshold.}
\end{table}

\subsubsection{Prevalence of Categories}\label{sec:res-pi2-prevalencia}  Table~\ref{tab:pi2} breaks down the eight RQ2 categories by stratum, given  H1\,=\,1. The PS column indicates the presence of substantial procedural content, which is not a mechanism of the conceptual framework but is an indicator that the rule file contains more than just an isolated rule.

\paragraph{M1 and M5 are near-universal; the rest are not.} Among those who already have a rule file, guiding the agent within the project (M1) and stating at least one rule (M5) are virtually the definition of what that file contains (Table~\ref{tab:pi2}) ---a result that is also consistent with the data in Table~\ref{tab:h5-idioma} (92.0--82.5\,\%), which was obtained by an independent lexical instrument on a different corpus: two distinct measurements of the same phenomenon converge. The remaining categories are much lighter: persistent knowledge with its rationale (M2, 10.0--26.7\,\%), executable specifications (M3, 10.0--20.0\,\%), prescribed structured consultation (M6, 5.0--20.0\,\%) and graduated autonomy  (M8, 13.3--26.7\,\%) ---representing the minimum and maximum across the four strata in Table~\ref{tab:pi2}. The narrow adoption shown in Table~\ref{tab:amplitud} is not just a phenomenon of \textit{who} adopts; it is also a phenomenon of \textit{how much} the adopted artifact contains. Even within the files that do exist, most of the harness described by the conceptual framework remains immaterialized.

\begin{table}[h]
\caption{Prevalence of each RQ2 category by stratum, conditioned on 
H1\,=\,1 (repositories with at least one rule file). PS =
\texttt{proc\_substantial}. Labels resolved through adjudication where coders disagreed.}\label{tab:pi2}
\footnotesize
\begin{tabular*}{\textwidth}{@{\extracolsep\fill}lrrrrrrrrr@{\extracolsep\fill}}
\toprule
Stratum & $n$ & M1 & M2 & M3 & M5 & M6 & M7 & M8 & PS \\
\midrule
P  & 60 & 93.3 & 21.7 & 10.0 & 95.0 & 5.0  & 66.7 & 13.3 & 25.0 \\
G1 & 30 & 100.0 & 26.7 & 16.7 & 93.3 & 6.7  & 90.0 & 13.3 & 43.3 \\
G2 & 30 & 93.3 & 26.7 & 20.0 & 96.7 & 20.0 & 56.7 & 26.7 & 56.7 \\
G3 & 30 & 96.7 & 10.0 & 13.3 & 86.7 & 13.3 & 53.3 & 20.0 & 16.7 \\
\botrule
\end{tabular*}
\footnotesize\emph{Source: src/qual/score\_rulefiles.py \texttt{->} data/pi2\_report\_it2.json.}
\end{table}

\paragraph{M6 here is not H6.} This M6 measures whether the rule file \textit{prescribes} that the agent must stop and ask; conversely, H6 in Section~\ref{sec:amenazas} measures whether that consultation was actually \textit{observed} within a PR comment. They represent two operationalizations of the same mechanism across two distinct corpora ---written policy versus \textit{enacted} behavior--- and should not be interpreted as the same measurement repeated. The fact that the H6 lexical classifier was discarded from the quantitative study due to low \textit{recall}  (Section~\ref{sec:amenazas}) does not affect this prevalence of M6: they rely on different instruments and corpora, and the human coder who assigns M6 does not depend on those lexical rules.

\paragraph{Substantive procedural content does not decrease with visibility, which constitutes an inductive finding.}  Unlike the remaining seven categories, \texttt{proc\_substantial} is not monotonic: G2 (56.7\,\%) exceeds  G1 (43.3\,\%), and both widely outperform P  (25.0\,\%) and G3 (16.7\,\%). This is the first empirical evidence—albeit still descriptive—of the confusion between procedural content and normative specification that Section~\ref{sec:metodo-cualitativo} explicitly codifies based on a distinction that is currently only conceptual (on-demand \textit{skills} versus always-loaded system specifications): the phenomenon exists, and it is not where the visibility gradient would predict it to be.

\subsubsection{Answer to RQ2}\label{sec:res-respuesta-rq2} The content of the rule files corresponds only partially to the taxonomy of the methodological harness. Agent guidance (M1) and normative specifications (M5) are near-universal, and evidence-based guidance (M7) is comparatively common. Conversely, shared persistent knowledge (M2), executable specifications (M3), structured consultation (M6), and graduated autonomy (M8) appear only in a minority of the files. The empirical artifacts therefore materialize selected components of the framework rather than the integrated methodological harness that it postulates.

\section{Threats to Validity and Limitations}\label{sec:amenazas}

\paragraph{Differential attrition in the lowest stratum.} In G3, one in seven repositories in the framework is no longer accessible, compared to one in thirty-seven in P. The figures for G3 are estimated based on survivors. If the repositories that disappear are disproportionately short---lived experiments and thus less harnessed—--G3 is overestimated, and the actual gradient would be more pronounced than the one observed. The bias would work against the finding rather than in its favor, making this a conservative interpretation; however, its direction cannot be verified with repositories that no longer exist.

\paragraph{Observed structured consultation (H6): the mechanism exists, but the lexical detector fails to capture it.} The validation of H6 against human coding (Section~\ref{sec:res-h6-validacion}) confirms that the mechanism is real and frequent: 53 confirmed consultations among the 536 comments sampled at full scale over stratum P, well above the threshold of ten positives fixed beforehand (Section~\ref{sec:metodo-cualitativo}), meaning that \textit{recall} is indeed estimable this time. However, the lexical classifier only retrieves 49.1\,\% of these confirmed consultations (precision 36.6\,\%), falling below the 0.70 threshold also fixed beforehand. A manual inspection of the false negatives and false positives explains why: most of the real consultations missed by the detector are formulated as prose questions with an implicit alternative (“should X be done, or would you prefer Y?”), without the explicit enumeration marker  (\texttt{1.}, \texttt{option A}) required by the rule; conversely, most of its false positives are numbered lists summarizing implementation states or steps within a PR description, rather than alternatives offered to a human. Based on an empirical design rule established a priori (Section~\ref{sec:metodo-cualitativo}), a real mechanism that the detector fails to detect more than half the time is not replaced by another method a posteriori: H6 is excluded from the quantitative study. The limitation is attributed to the instrument, not to the mechanism.

\paragraph{False negatives by convention.} The detectors are bound to the file names currently in use: any team implementing a mechanism under a different convention remains invisible, meaning that all prevalence rates in this study represent lower bounds. The most relevant case is persistent knowledge (H2), whose figure is near-zero (1.9\,\% in the most visible stratum (P)), with intervals that fully overlap among the three remaining strata: there is no resolvable gradient because there is virtually nothing to scale. These data do not distinguish whether this practice does not exist or if it is materialized where a public repository does not expose it—be it in the tool's memory, internal documentation, or the history of conversations with the agent.


\paragraph{A zero can represent either absence or invisibility: the case of H2.} Furthermore, there is a design reason, not just a conventional one, why this margin is larger for H2 than for the other mechanisms. The framework \citep{diaz2026sdd} defines it as an a\textit{gent- and tool-agnostic} store: a team asset, rather than the private memory of a single harness. When the mechanism is materialized according to this definition, it tends to do so as an external service to which any agent connects via a protocol (for instance, dedicated MCP servers designed to persist decisions and context across sessions, such as Engram \citep{engram2026}, Mem0 \citep{mem02026} or Cipher \citep{cipher2026}). By construction, this design places them outside the versioned tree and the Git history that this study can read, precisely to avoid leaving a footprint in it. Adopting these tools does not leave the trace that H2 looks for, even if the practice that H2 aims to capture is present. The near-absence observed might therefore not be a merely contingent limitation of the instrument: it could be, in part, exactly what is to be expected when the mechanism is adopted in the manner prescribed by the framework itself. Distinguishing absence from invisibility requires working with practitioners, which could be scheduled as a new study..

\paragraph{Early adopters.} The sample frame consists of repositories with agentic activity prior to August 2025. If those who adopt agents early also adopt the harness early, the prevalences estimated here overestimate those of the general population of repositories with agentic activity. The reweighted estimate of 21.7\,\% corrects for \textit{popularity} bias, which is what the stratified design allows us to correct; it does not correct for \textit{early adoption} bias, which remains.

\paragraph{No process results.} This study does not assess whether the harness works. No outcome variables have been measured, and no claims regarding efficacy can be drawn from these figures.

\section{Discussion}\label{sec:discusion}

\subsection{RQ1: The harness is observable, but widespread adoption is uncommon}\label{sec:disc-rq1-adopcion}

RQ1 shows that harness adoption is \textit{narrow and uneven}: one mechanism dominates while the others are marginal, and adopting repositories materialize one or two mechanisms rather than a coherent whole.

Two-thirds of the most visible repositories incorporate \textit{something}, but the mean breadth is 1.15 mechanisms out of five, and a mere 2.6\,\% of that stratum combines four or five mechanisms. Outside of it, this figure drops to three repositories in G1, five in G2, and \textit{none} in G3. Reweighted to the population, only 21.7\,\% of repositories with agentic activity possess any harness mechanism.

Furthermore, what is adopted is not distributed at random. The frequent combinations consist of rule files accompanied by governance artifacts that many projects already possessed—--such as PR templates and \texttt{CODEOWNERS}--—whereas combinations involving strictly methodological mechanisms are rare: executable specifications alongside rule files appear in 182 out of 5,435 repositories, and persistent knowledge alongside rule files in 66. The interpretation we propose is that \textbf{what has been predominantly adopted is an artifact, rather than a method}. Writing a \texttt{CLAUDE.md} file carries the cost of a single afternoon; sustaining a decision record or a body of executable specifications requires a continuous methodological investment, and that is exactly the part that does not occur
.

\subsection{RQ1: Adoption varies heavily with repository visibility}\label{sec:disc-rq1-gradiente}

The cleanest result of the study is that all five mechanisms decrease monotonically with visibility, without a single exception.

However, the order in which they drop reveals something that the conceptual framework did not anticipate. The two mechanisms that drop the most (P/G3 row, Table~\ref{tab:prevalencia}) ---\texttt{CODEOWNERS} ($\times$8.7) and PR templates ($\times$6.2)--- are precisely the two that \textbf{presuppose a team}: assigning reviewers by path and structuring communication between the contributor and the reviewer make no sense for a single person. The mechanism that holds up best is rule files ($\times$4.1), which an individual developer writes in the exact same way as a team does, because it is directed at the agent rather than at other humans.

It is therefore possible that a substantive part of the visibility gradient is a gradient of \textit{team presence}. This is an interpretation, not a measurement: validating it requires a direct indicator of team size that this study does not incorporate. However, it carries an immediate consequence for the framework. If some of the postulated mechanisms are actually human coordination artifacts that predate the agentic era---\texttt{CODEOWNERS} existed long before agents--—then the harness, as operationalized here, mixes two things: devices created \textit{to govern} agents and pre-existing devices \textit{repurposed} for that reason. The distance between the two temporal estimands points in the same direction and quantifies it: there is empirical evidence that the adoption of purely agentic mechanisms doubles between the observed work and cloning (reaching a ratio of $\approx \times 2.1$), whereas pre-existing team infrastructure remains virtually static with a growth ratio of only $\times 1.18$.

It is also the interpretation most consistent with the classic observation that the structure of a system’s artifacts reflects the structure of the organization that produces them  \citep{conway1968committees}: where there is no organization to reflect, coordination artifacts do not appear, whether agents are present or not.

\subsection{RQ1: Mechanisms specific to the agentic era show substantial recent adoption}\label{sec:disc-rq1-temporal}

The two temporal estimands separate what the conceptual framework does not. Evaluated at the moment each PR was created, the mechanisms specific to the agentic era—--rule files, persistent knowledge, and executable specifications—--\textbf{approximately doubled} their presence between the observed work and the data collection date ($\times$2.00 a $\times$2.12), whereas the two pre-existing governance artifacts barely moved ($\times$1.18). They were already there.

Harness growth is therefore not uniform: it is concentrated precisely in the components created as a direct response to the introduction of agents. This constitutes the most direct evidence from this study in favor of the framework’s core thesis--—that the arrival of agents compels the construction of something new around the team--—and it is worth noting that this insight stems from a comparison made possible only because first-appearance dates were measured, rather than mere presence.

While the temporal analysis supports a conclusion regarding the level of recent adoption, it does not do so for its acceleration rate. The level is clear: 58.3\,\% of rule files appeared after the window closure. However, \textit{acceleration} is not. The monthly histogram of first appearances peaks in June 2025 and declines thereafter, but this cannot be interpreted as a cooling off: first appearances are being counted, a repository can only appear once, and the raw count necessarily decays as the pool of repositories that have not yet adopted is exhausted. Claiming acceleration requires evaluating cumulative incidence over the population at risk, an analysis that remains as future work. In this context, the population at risk comprises only those repositories with agentic activity that have not yet adopted a rule file by the observed month—a pool that inevitably shrinks over time. What is indeed interpretable, being far removed from both this exhaustion effect and the selection bias of the sampling frame, is the rebound in early 2026: adoption did not halt with the period that defined the sample.

\subsection{RQ2: The content of the rule files corresponds only partially to the framework}\label{sec:disc-rq2}

The results from RQ2 (Section~\ref{sec:res-pi2}) prove the same thesis of narrow adoption from a different angle and reinforce it. Instead of asking \textit{who} has a rule file, it asks \textit{what is inside} those that already exist, and the answer reveals the exact same asymmetry in miniature: guiding the agent (M1) and stating at least one rule (M5) are near-universal, whereas sustaining shared persistent knowledge with its rationale, specifying behavior in a verifiable manner, prescribing structured consultation, or graduating autonomy appear only in a minority of the corpus—even when conditioned on the file's pre-existence. \textbf{Narrow adoption is therefore not just a question of how many repositories adopt something; it is also a question of how much of that something actually gets written}, and both questions point independently in the exact same direction.

The most unexpected inductive finding of RQ2 points another way. Substantive procedural content housed within always-loaded files—the confusion between on-demand \textit{skills} and normative specifications that the framework distinguishes only conceptually—does not decrease with visibility like the other mechanisms: G2 (56.7\,\%) and G1 (43.3\,\%) widely outperform P (25.0\,\%) and G3 (16.7\,\%). The phenomenon exists, and it does not reside where the visibility gradient would lead one to assume. Any revision of the framework that treats this confusion as a problem characteristic of less mature projects would be mistaken in its direction.

\subsection{Implications of the Two Research Questions for the Framework}\label{sec:disc-implicaciones}

Together, RQ1 and RQ2 qualify the framework without refuting it. \citet{diaz2026sdd} describe the harness as a whole whose mechanisms reinforce one another; however, the data show that while these mechanisms exist individually, such an integrated whole is currently an exceptional configuration. The framework therefore describes a goal rather than a widespread practice, and this distinction must be made explicit when used prescriptively.

Furthermore, the evidence suggests two specific revisions. First, to explicitly distinguish mechanisms created to govern agents from pre-existing coordination artifacts repurposed for that end (Section~\ref{sec:disc-rq1-gradiente}); the temporal data separate them cleanly, whereas the framework treats them identically. Second, to treat separately whether a mechanism is \textit{prescribed}, i.e., written as a policy in the rule file, and whether it is enacted, i.e., observed as behavior in a public artifact. The current framework blurs both questions under the same mechanism; RQ2 measures solely the prescribed side through M6 (5.0–20.0\,\% depending on the stratum, conditioned on having a rule file). This study's attempt to also measure the enacted side through PR comments proved inconclusive  (Section~\ref{sec:amenazas}). This precisely illustrates why the distinction matters: without it, a mechanism is declared absent when, in reality, it is the observation channel that fails.

For research, the difference between the 65.3\,\% of the \textit{popular} stratum and the reweighted 21.7\,\% is not a methodological nuance; it is a factor of three. A study applying the customary filter—more than one hundred stars—would have correctly described a population of 2,807 repositories and presented it as if it described 116,211. We recommend that studies on agentic practice adoption explicitly declare the population their figures address and, when costs permit, measure the gradient instead of assuming it to be negligible.

For practice, the majority configuration among adopters consists of a rule file accompanied by pre-existing governance artifacts. If the framework's thesis is correct—and this study does not test it—then most of the expected benefit stems from the mechanisms that almost no one has adopted. A team that has written its \texttt{CLAUDE.md} and considers the transition complete is, according to these data, in the modal configuration, not the advanced one.

\section{Conclusions}\label{sec:conclusiones}
The initial question was whether the harness postulated by \citet{diaz2026sdd} exists in practice, and to what extent. The answer supported by these data consists of three parts, and none of the three is a simple yes or no.

\textbf{It exists, but as an artifact rather than as a whole}. Reweighted to the 116,211 repositories with agentic activity in AIDev, the adoption of any harness mechanism stands at 21.7\,\%, and combining four or five out of the five common mechanisms is exceptional even in the most visible stratum. What is predominantly adopted is a rule file accompanied by governance artifacts that many projects already possessed before the arrival of agents, rather than the mutually reinforcing whole described by the framework. RQ2 reaches the same conclusion through a different path: even within the files that do exist, guiding the agent and stating at least one rule are near-universal, whereas persistent knowledge, executable specifications, structured consultation, and graduated autonomy appear only in a minority of the corpus. Narrow adoption is therefore not just a question of who adopts; it is also a question of how much of the adopted artifact actually gets written.

\textbf{Where it exists, it behaves as would be expected from a recent and uneven phenomenon}. The visibility gradient is monotonic without exception across all five common mechanisms, and its shape suggests that it substantively measures team presence rather than harness adoption: mechanisms that presuppose human coordination drop more sharply than those an individual developer can write alone. The majority of the observed rule files appeared after the closure of the agentic work window that defines the sample, and growth was disproportionately concentrated in the components specifically created to govern agents, rather than in pre-existing governance artifacts repurposed for that end.

\textbf{And where it cannot be established, it is explicitly stated rather than filled with assumptions}. The language-based coverage of the normative content detector partially confounds the gradient it measures, and this effect is quantified rather than merely declared. The automated detection of structured consultation (H6) in PR comments confirmed that the mechanism is real (53 hand-confirmed cases), but the lexical classifier only retrieves half of them; the mechanism is discarded from the quantitative study instead of interpreting this insufficient recall as if the mechanism barely existed, which would have been the opposite error.

The practical takeaway is straightforward to state and easy to overlook: \textbf{a team that has written its CLAUDE.md and considers the transition toward agentic software engineering complete is, according to these data, in the modal configuration, not the advanced one}. If the framework’s thesis is correct—and this study tests it but does not validate it in its entirety—most of the expected benefit from the harness stems precisely from the mechanisms that almost no one has adopted yet.

This study establishes presence, not effect. This is a deliberate limitation, not an omission: asking whether the harness works before knowing if it exists would have been building on sand. With presence now measured and its instrument published, the question of effect can be raised on firm ground in future work
.

\backmatter

\bmhead{Acknowledgments}
GenAI tools were used in a strictly assistive role to support English language editing, instruments' programming, and to suggest local revisions of the manuscript. All scientific content, methodological decisions, analysis and interpretation of results are the sole responsibility of the authors, who reviewed and validated every AI-assisted edit.

\section*{Declarations}

\bmhead{Funding} Not applicable.
\bmhead{Conflict of interest} The authors declare that they have no conflict of interest.
\bmhead{Data availability}\label{sec:data-availability} All the data related to this research, providing a detailed chain of evidence, is available at https://github.com/jdiazfernandez/empiricalstudy-ase-mining-swrepositories. For assuring long-term archiving, the data is also stored https://doi.org/10.5281/zenodo.22885512.

\bibliography{references}

@misc{diaz2026sdd,
  author        = {Jessica D{\'\i}az and Joaquin Gayoso and Andrea Cimminio and Jorge P{\'e}rez},
  title         = {Spec-Driven Development for Agentic Software Engineering: Harnessing Human Agent Teamwork},
  year          = {2026},
  eprint        = {},
  archivePrefix = {arXiv},
  primaryClass  = {cs.SE},
  url           = {}
}

@misc{aidev2025dataset,
  author       = {Li, Hao and Zhang, Haoxiang and Hassan, Ahmed E.},
  title        = {{AIDev}: A Dataset of Agentic Pull Requests},
  year         = {2025},
  howpublished = {Hugging Face Datasets},
  url          = {https://huggingface.co/datasets/hao-li/AIDev},
}

@article{kalliamvakou2016perils,
  author  = {Kalliamvakou, Eirini and Gousios, Georgios and Blincoe, Kelly and
             Singer, Leif and German, Daniel M. and Damian, Daniela},
  title   = {An in-depth study of the promises and perils of mining {GitHub}},
  journal = {Empirical Software Engineering},
  volume  = {21},
  number  = {5},
  pages   = {2035--2071},
  year    = {2016},
  doi     = {10.1007/s10664-015-9393-5}
}

@article{munaiah2017curating,
  author  = {Munaiah, Nuthan and Kroh, Steven and Cabrey, Craig and Nagappan, Meiyappan},
  title   = {Curating {GitHub} for engineered software projects},
  journal = {Empirical Software Engineering},
  volume  = {22},
  number  = {6},
  pages   = {3219--3253},
  year    = {2017},
  doi     = {10.1007/s10664-017-9512-6}
}

@article{cohen1960kappa,
  author  = {Cohen, Jacob},
  title   = {A coefficient of agreement for nominal scales},
  journal = {Educational and Psychological Measurement},
  volume  = {20},
  number  = {1},
  pages   = {37--46},
  year    = {1960},
  doi     = {10.1177/001316446002000104}
}

@article{landis1977agreement,
  author  = {Landis, J. Richard and Koch, Gary G.},
  title   = {The measurement of observer agreement for categorical data},
  journal = {Biometrics},
  volume  = {33},
  number  = {1},
  pages   = {159--174},
  year    = {1977},
  doi     = {10.2307/2529310}
}

@article{nosek2018preregistration,
  author  = {Nosek, Brian A. and Ebersole, Charles R. and DeHaven, Alexander C. and
             Mellor, David T.},
  title   = {The preregistration revolution},
  journal = {Proceedings of the National Academy of Sciences},
  volume  = {115},
  number  = {11},
  pages   = {2600--2606},
  year    = {2018},
  doi     = {10.1073/pnas.1708274114}
}

@article{GAROUSI:2019,
title = {Guidelines for including grey literature and conducting multivocal literature reviews in software engineering},
journal = {Information and Software Technology},
volume = {106},
pages = {101-121},
year = {2019},
issn = {0950-5849},
doi = {https://doi.org/10.1016/j.infsof.2018.09.006},
url = {https://www.sciencedirect.com/science/article/pii/S0950584918301939},
author = {Vahid Garousi and Michael Felderer and Mika V. Mäntylä}
}

@misc{fowler2026harness,
  author       = {B\"{o}ckeler, Birgitta and Fowler, Martin},
  title        = {Harness Engineering for Coding Agent Users},
  year         = {2026},
  howpublished = {martinfowler.com},
  url          = {https://martinfowler.com/articles/harness-engineering.html},
  note         = {Practitioner article. Accessed June 2026}
}

@misc{langchain2026harness,
  author       = {{LangChain}},
  title        = {The Anatomy of an Agent Harness},
  year         = {2026},
  howpublished = {LangChain blog},
  url          = {https://www.langchain.com/blog/the-anatomy-of-an-agent-harness},
  note         = {Practitioner article. Accessed June 2026}
}

@misc{claudecode2026worktrees,
  author       = {{Anthropic}},
  title        = {Run Parallel Sessions with Worktrees},
  year         = {2026},
  howpublished = {Claude Code Docs},
  url          = {https://code.claude.com/docs/en/worktrees},
  note         = {Official documentation. Accessed August 2026}
}

@misc{openai2026codexcloud,
  author       = {{OpenAI}},
  title        = {Codex Cloud},
  year         = {2026},
  howpublished = {ChatGPT Learn, developers.openai.com},
  url          = {https://developers.openai.com/codex/cloud},
  note         = {Official documentation. Accessed August 2026}
}

@misc{opencode2026agents,
  author       = {{OpenCode}},
  title        = {Agents},
  year         = {2026},
  howpublished = {OpenCode Docs},
  url          = {https://opencode.ai/docs/agents/},
  note         = {Official documentation. Accessed August 2026}
}

@misc{engram2026,
  author       = {{Gentleman-Programming}},
  title        = {Engram: Persistent Memory for {AI} Coding Agents},
  year         = {2026},
  howpublished = {GitHub repository},
  url          = {https://github.com/Gentleman-Programming/engram},
  note         = {Agent-agnostic MCP memory server. Accessed August 2026}
}

@misc{mem02026,
  author       = {{Mem0AI}},
  title        = {Mem0: Universal Memory Layer for {AI} Agents},
  year         = {2026},
  howpublished = {GitHub repository},
  url          = {https://github.com/mem0ai/mem0},
  note         = {Open-source persistent memory {SDK} for {AI} agents. Accessed August 2026}
}

@misc{cipher2026,
  author       = {{Byterover}},
  title        = {Cipher: Memory Layer for Coding Agents},
  year         = {2026},
  howpublished = {GitHub repository},
  url          = {https://github.com/campfirein/cipher},
  note         = {Renamed {ByteRover CLI} after this study's corpus was mined. Accessed August 2026}
}

@techreport{farosai2025paradox,
  author      = {{Faros AI}},
  title       = {The {AI} Productivity Paradox Report},
  institution = {Faros AI},
  year        = {2025},
  url         = {https://www.faros.ai/blog/ai-software-engineering},
  note        = {Industry report. Accessed June 2026}
}

@techreport{dora2024,
  author      = {{Google Cloud / DORA}},
  title       = {Accelerate State of {DevOps} Report 2024},
  institution = {Google Cloud},
  year        = {2024},
  url         = {https://dora.dev/research/2024/dora-report/}
}

@techreport{dora2025,
  author      = {{Google Cloud / DORA}},
  title       = {State of {AI}-assisted Software Development},
  institution = {Google Cloud},
  year        = {2025},
  url         = {https://dora.dev/research/2025/dora-report/}

}

@techreport{metr2025,
  author      = {{METR}},
  title       = {Measuring the Impact of Early-2025 {AI} Tools on Experienced Open-Source Developer Productivity},
  institution = {METR},
  year        = {2025},
  url         = {https://metr.org/blog/2025-07-10-early-2025-ai-experienced-os-dev-study/}
}

@article{he2025llmagents,
  author  = {He, Junda and Treude, Christoph and Lo, David},
  title   = {{LLM}-Based Multi-Agent Systems for Software Engineering: Literature Review, Vision, and the Road Ahead},
  journal = {ACM Transactions on Software Engineering and Methodology},
  volume  = {34},
  number  = {5},
  year    = {2025},
  doi     = {10.1145/3712003}
}

@article{amalfitano2026,
author = {Amalfitano, Domenico and Metzger, Andreas and Autili, Marco and Fulcini, Tommaso and Hey, Tobias and Keim, Jan and Pelliccione, Patrizio and Scotti, Vincenzo and Koziolek, Anne and Mirandola, Raffaela and Vogelsang, Andreas},
title = {A Research Roadmap for Augmenting Software Engineering Processes and Software Products with Generative AI},
year = {2026},
publisher = {Association for Computing Machinery},
address = {New York, NY, USA},
issn = {1049-331X},
url = {https://doi.org/10.1145/3788879},
doi = {10.1145/3788879},
note = {Just Accepted},
journal = {ACM Trans. Softw. Eng. Methodol.},
month = jan
}

@article{ahmed2025,
author = {Ahmed, Iftekhar and Aleti, Aldeida and Cai, Haipeng and Chatzigeorgiou, Alexander and He, Pinjia and Hu, Xing and Pezz{\`e}, Mauro and Poshyvanyk, Denys and Xia, Xin},
title = {Artificial Intelligence for Software Engineering: The Journey So Far and the Road Ahead},
year = {2025},
issue_date = {June 2025},
publisher = {Association for Computing Machinery},
address = {New York, NY, USA},
volume = {34},
number = {5},
issn = {1049-331X},
url = {https://doi.org/10.1145/3719006},
doi = {10.1145/3719006},
journal = {ACM Trans. Softw. Eng. Methodol.},
month = may,
articleno = {119},
numpages = {27}
}

@article{liu2026,
author = {Liu, Junwei and Wang, Kaixin and Chen, Yixuan and Peng, Xin and Chen, Zhenpeng and Zhang, Lingming and Lou, Yiling},
title = {Large Language Model-Based Agents for Software Engineering: A Survey},
year = {2026},
publisher = {Association for Computing Machinery},
address = {New York, NY, USA},
issn = {1049-331X},
url = {https://doi.org/10.1145/3796507},
doi = {10.1145/3796507},
note = {Just Accepted},
journal = {ACM Trans. Softw. Eng. Methodol.},
month = mar
}

@inproceedings{li2026teammates,
author = {Li, Hao and Zhang, Haoxiang and Zhang, Jie M. and Lou, Yiling and Ray, Baishakhi and Zimmermann, Thomas and Hassan, Ahmed E.},
title = {Agentic Software Engineering (SE 3.0): The Rise of AI Teammates},
year = {2026},
isbn = {9798400722592},
publisher = {Association for Computing Machinery},
address = {New York, NY, USA},
url = {https://doi.org/10.1145/3770855.3818242},
doi = {10.1145/3770855.3818242},
booktitle = {Proceedings of the 32nd ACM SIGKDD Conference on Knowledge Discovery and Data Mining V.2},
pages = {13417–13418},
numpages = {2},
location = {Republic of Korea},
series = {KDD '26}
}

@article{conway1968committees,
  title={How do committees invent?},
  author={Conway, Melvin E.},
  journal={Datamation},
  volume={14},
  number={4},
  pages={28--31},
  year={1968}
}

\begin{appendices}
\renewcommand{\thetable}{\Alph{section}.\arabic{table}}%
\renewcommand{\thefigure}{\Alph{section}.\arabic{figure}}%
\makeatletter\@addtoreset{table}{section}\@addtoreset{figure}{section}\makeatother

\section{AIDev Scale by Stratum}\label{sec:apx-aidev}

Table~\ref{tab:apx-aidev} summarizes, by stratum, the volume of AIDev data supporting this study: selected and successfully cloned repositories, and agentic pull requests (PRs).

\begin{table}[h]
\caption{AIDev scale by stratum. ``Agentic PRs'' is derived directly from AIDev for the selected repositories, regardless of whether the local cloning was
successful.}\label{tab:apx-aidev}
\footnotesize
\begin{tabular*}{\textwidth}{@{\extracolsep\fill}lrrr@{\extracolsep\fill}}
\toprule
Stratum & Repositories (sample) & Successfully cloned & Agentic PRs \\
\midrule
P     & 2,807 & 2,732 & 33,596 \\
G1    & 1,000 & 965   & 7,299  \\
G2    & 1,000 & 882   & 9,910  \\
G3    & 1,000 & 856   & 8,897  \\
\midrule
Total & 5,807 & 5,435 & 59,702 \\
\botrule
\end{tabular*}
\footnotesize\emph{Source: data/sample\_frame.csv, data/step\_zero\_report.json,
data/harness\_\{P,G1,G2,G3\}\_raw.json.}
\end{table}

AIDev only distributes PR comments, reviews, and review comments for the \textit{popular} tier (stratum P); they do not exist for the rest of the population  (Section~\ref{sec:metodo-datos}), which is why H6 (structured consultation) is observable only in P (Section~\ref{sec:metodo-mecanismos}). The H6 detector (Table~\ref{tab:h6-cajones}) reads three text surfaces across the 33,596 agentic PRs in P --- the body of each PR, its comments, and its review comments --- as a single corpus of 99,225 text units; Table~\ref{tab:apx-comentarios} gives the composition. The \texttt{pr\_reviews} table is used only to link each review comment to its PR and contributes no text of its own.

\begin{table}[h]
\caption{Composition of the H6 text corpus, stratum P (the only stratum for which AIDev distributes comment tables), by surface and authorship. PR bodies are all agent-authored because every PR in AIDev was opened by an agent. Rows with a null body are excluded (one review comment).}\label{tab:apx-comentarios}
\footnotesize
\begin{tabular*}{\textwidth}{@{\extracolsep\fill}lrrr@{\extracolsep\fill}}
\toprule
Text surface (AIDev table) & Agent-authored & Human-authored & Total \\
\midrule
PR bodies (\texttt{pull\_request})                      & 33,236 & 0      & 33,236 \\
PR comments (\texttt{pr\_comments})                     & 14,145 & 24,977 & 39,122 \\
Review comments (\texttt{pr\_review\_comments\_v2})     & 9,178  & 17,689 & 26,867 \\
\midrule
Total                                                   & 56,559 & 42,666 & 99,225 \\
\botrule
\end{tabular*}
\footnotesize\emph{Source: data/consultations\_full.parquet (src/classify/consultation\_full.py).}
\end{table}

\section{Unit of Analysis in Prevalence, E1 and E2}\label{sec:apx-e2}

  The figures in Table~\ref{tab:prevalencia} and Table~\ref{tab:e2} are not directly comparable because they differ across two independent axes of the unit of analysis: the applied estimand and the weighting system used. This appendix makes both axes explicit before presenting, in Table~\ref{tab:apx-e2-comparacion}, how the exact same question yields three distinct figures depending on which axis is fixed.

\begin{itemize}
  \item \textbf{Which estimand is applied.} E1 asks whether the mechanism exists at the time of cloning (in the repository's \texttt{HEAD}); E2 asks whether it already existed at the exact moment a specific PR was created (Section~\ref{sec:metodo-estimandos}).
  \item \textbf{What is weighted.} Each repository can count once (weighting by repository), or each agentic PR can count once (weighting by PR), meaning that a repository with 500 PRs carries 500 times more weight than one with only a single PR.
\end{itemize}

These two axes are independent, and only three out of the four possible combinations appear in the article. E2 only makes sense when weighted by PR: ''the exact moment this PR was created'' does not exist at the repository level, only at the PR level. In order to compare E2 against E1 within the same table (Table~\ref{tab:e2}), E1 is also recalculated using PR-based weighting—--applying the exact same definition of E1 used in  Table~\ref{tab:prevalencia} (which is weighted by repository), but under a different weighting system. Table~\ref{tab:apx-e2-comparacion} summarizes this for H1 as an example; the pattern remains identical for the other mechanisms.

\begin{table}[h]
\caption{The exact same question (Does the rule file, H1, exist?) interpreted in three distinct ways depending on the estimand and the weighting system. None of these figures is incorrect; each answers a different question.}\label{tab:apx-e2-comparacion}
\footnotesize
\begin{tabular*}{\textwidth}{@{\extracolsep\fill}lllr@{\extracolsep\fill}}
\toprule
 & Estimand & Weighting & H1 in P \\
\midrule
Prevalence (Table~\ref{tab:prevalencia})    & E1 (exists at cloning)              & By repository & 51.2\,\% \\
E1 Column  (Table~\ref{tab:e2})              & E1 (exists at cloning)              & By PR          & 55.7\,\% \\
E2 Column (Table~\ref{tab:e2})              & E2 (existed when opening the PR)  & By PR          & 26.4\,\% \\
\botrule
\end{tabular*}
\footnotesize\emph{Source: data/prevalence\_report.json (fila 1) y
data/coverage\_e2\_report.json (filas 2-3).}
\end{table}

The fact that the first and second rows differ (51.2\,\% versus 55.7\,\%) is not a calculation error: they represent the exact same definition of E1 under two different weighting systems, and the difference between them indicates that repositories with higher agentic activity adopt H1 somewhat more frequently than average, thereby driving the weight upward when counting by PR rather than by repository. On the other hand, the fact that the third row is substantially lower (26.4\,\%) is indeed substantive: it demonstrates that a large portion of the adoption recorded by E1 occurred after the agentic work being counted, rather than before it  (Section~\ref{sec:res-e2}).

\end{appendices}

\end{document}